\documentclass[aps,prd,preprintnumbers,showpacs,nofootinbibn,twocolumn]{revtex4-1}

\usepackage{overpic}
\usepackage{subfigure}
\usepackage[utf8]{inputenc}
\usepackage[english]{babel}
\usepackage{amsmath}
\usepackage{amsfonts}
\usepackage{amssymb}
\usepackage{epsfig}
\usepackage{graphics,psfrag,rotating}
\usepackage{graphicx}
\usepackage{dcolumn}
\usepackage{bm}

\begin{document}

\newcommand{\Od}{{\cal O}}
\newcommand{\lsim}   {\mathrel{\mathop{\kern 0pt \rlap
  {\raise.2ex\hbox{$<$}}}
  \lower.9ex\hbox{\kern-.190em $\sim$}}}
\newcommand{\gsim}   {\mathrel{\mathop{\kern 0pt \rlap
  {\raise.2ex\hbox{$>$}}}
  \lower.9ex\hbox{\kern-.190em $\sim$}}}


\title{Kerr-Newman black holes in $f(R)$ theories}
\thispagestyle{empty}

\author{J.\,A.\,R.\,Cembranos$\,^{(a)}$\footnote{E-mail: cembra@fis.ucm.es},
A.\,de la Cruz-Dombriz$\,^{(b,c)}$\footnote{E-mail: alvaro.delacruz-dombriz@uct.ac.za}
and P. Jimeno Romero$\,^{(a)}$\footnote{E-mail: jimeno.romero@gmail.com}
}

\address{$^{(a)}$ Departamento de F\'{\i}sica Te\'orica I, Universidad Complutense de Madrid, E-28040 Madrid, Spain.}
\address{$^{(b)}$ Astrophysics, Cosmology and Gravity Centre (ACGC), University of Cape Town, Rondebosch, 7701, South Africa}
\address{$^{(c)}$ Department of Mathematics and Applied Mathematics, University of Cape Town, 7701 Rondebosch, Cape Town, South Africa.}


\date{\today}


\begin{abstract}

In the context of $f(R)$ modified gravity theories, we study the Kerr-Newman black-hole solutions.
We study non-zero constant scalar curvature solutions and discuss the metric tensor that satisfies the modified field equations. We determine that,  in absence of a cosmological constant, the black holes existence is determined by the sign of a parameter dependent of the mass, the charge, the spin and the scalar curvature. We obtain that for negative values of the curvature, the \textnormal{extremal} black hole is no longer given by a spin parameter $a_{max}=M$ (as is the case in General Relativity), but by $a_{max}<M$, and that for positive values of the curvature there are two kinds of extremal black holes: the usual one, that occurs for $a_{max}>M$, and the \textnormal{extreme marginal} one, where the exterior (but not interior) black hole's horizon vanishes provided that $a<a_{min}$. Thermodynamics for this kind of black holes is then studied, as well as their local and global stability. Finally we study different $f(R)$ models and see how these properties
manifest for their parameters phase space.
\end{abstract}

\maketitle

\section{Introduction}

General Relativity (GR) has been the most successful gravitational theory of the last century, fully accepted as a theory that describes the macroscopic geometrical properties of spacetime. For an isotropic and homogeneous geometry, GR leads to Friedmann equations which describe in an appropriate way the cosmological evolution with radiation and then matter dominated epochs. Nevertheless, the development of observational cosmology in the last decades with experiments of increasing precision like supernovae observations \cite{SIa}, has revealed that the Universe is in a stage of accelerated expansion. GR provided with usual matter sources is not able to explain this phenomenon. Moreover, GR does not account either for the cosmological era known as inflation \cite{inflac}, believed to have taken place before the radiation stage and that could alleviate
some problems of standard cosmology like the horizon and the flatness problem \cite{Peebles}. In addition, GR with usual baryonic matter cannot explain the observed matter density 
determined by fitting the standard $\Lambda\text{CDM}$ model to the WMAP7 data (Wilkinson Microwave Anisotropy Probe results for 7 years of observations) \cite{Komatsu}, the latest measurements from the BAO (Baryon Acoustic Oscillations) in the distribution of galaxies \cite{Percival}
and the Hubble constant ($H_0$) measurement \cite{Riess}. Thus, GR requires the introduction of an extra component called dark matter (DM), that accounts for about $20\%$ of the energy content of our Universe. Although there are many possible origins for this component \cite{DM}, DM is usually assumed to be in the form of thermal relics that naturally freeze-out with the right abundance in many extensions of the standard model of particles \cite{WIMPs}. Future experiments will be able to discriminate among the large number of candidates and model, such as direct and indirect detection designed explicitly for their search \cite{isearches}, or even at high energy colliders, where they could be produced \cite{Coll}.

A more puzzling problem is associated to the present accelerated expansion of the Universe. There are also a large amount of different explanations. One of them, assuming the validity of GR, postulates the existence of an extra cosmic fluid, the dark energy (DE), whose state equation $p=\omega_{\text{DE}} \rho$ (where $p$ and $\rho$ are the pressure and the energy density of the fluid) demand $\omega_{\text{DE}}<-1/3$ in order to provide an accelerated cosmic expansion \cite{DE}. The cosmological constant is the simplest model of DE, corresponding to an equation of state $\omega_{\text{DE}}=-1$. However, if we assume that the cosmological constant represents the quantum vacuum energy, its value seems to be many orders of magnitude bigger than the observed one \cite{cosmoproblema}.

In addition, there is also the problem that, as this cosmological constant cannot account for an inflationary period, a slow rolling scalar field, the {\it inflaton}, has to be introduced by hand. Nevertheless, other explanations for the mentioned acceleration may be provided by theories
that modify GR by considering actions different from the Einstein-Hilbert one \cite{varios}. Examples are Lovelock theories, free of ghosts and whose field equations contain second derivatives of the metric at most; string theory inspired models, that include a Gauss-Bonnet term in the Lagrangian; scalar-tensor theories like Brans-Dicke one, in which gravitational interaction is mediated by both a scalar field and GR tensor field; or the so called $f(R)$ theories, in which our work will be focused.  In this work we shall restrict ourselves to $f(R)$ theories in the metric formalism (where the connection depends on the metric, so the present fields in the gravitational sector of the action come only from the metric tensor) in the Jordan frame. In this frame, the gravitational Lagrangian is given by $R+f(R)$, where $f(R)$ is an arbitrary function of the scalar curvature $R$, and Einstein's equations usually become of fourth order on the metric derivatives.

$f(R)$ theories were proved (see \cite{Odintsov} among others) to be able to mimic the whole cosmological history, from inflation to the actual accelerated expansion era.
Diverse applications of these theories on gravitation and cosmology have been also widely studied  \cite{Tsujikawa}, as well as multiple ways to observationally and experimentally distinguish them from GR.  Concerning local tests of gravity and other cosmological constraints, see \cite{varia}.

The study of alternative gravitational theories to GR requires to confirm or discard their validity by obtaining solutions that can describe correctly, e.g., the cosmological evolution, the  growth factor of  cosmological perturbations and the existence of GR-predicted astrophysical objects such as black holes (BH). It is a well-known fact that, by choosing an appropriate function, $f(R)$ theories can mimic any cosmological evolution and, in particular, the one described by the $\Lambda$CDM model \cite{mimic_lambdacdm}. In fact, some modified gravity theories present the so called {\it degeneracy problem}: from large scale observations (Ia type supernova, BAO, or the cosmic microwave background) which depend uniquely on the evolution history of the Universe, the nature and the origin of DE cannot be determined due to the fact that identical evolutions can be explained by a diverse number of theories. However, it has been proved \cite{Dombriz_perturbaciones_PRD} that when scalar cosmological perturbations are studied, $f(R)$ theories, even mimicking the standard cosmological expansion, provide a different matter power spectrum from that predicted by the $\Lambda$CDM model \cite{Comment}.
Therefore, it is interesting to study the properties of BH in this kind of theories, since some of their known features might be either exclusive of Einstein's gravity or intrinsic features of any covariant gravitational theory. On the other hand, obtained results could provide a method to discard models  that disagree with expected physical results. In this sense research of BH thermodynamics may shed some light about the viability of alternative gravity theories since local and global stability regions, and consequently the existence itself of BH, depend on the values of the parameters of the model under consideration.


BH properties have been widely studied in other modified gravity theories: for instance \cite{cvetic,Cai_GaussBonet_AdS} studied BH in Einstein's theory with a Gauss-Bonnet term and a cosmological constant. Gauss-Bonnet and/or quadratic Riemann interaction terms are studied in \cite{Cho}, where is found that for a negative curvature of the horizon, phase transitions might occur. BH in Lovelock gravitational theories were studied in \cite{Matyjasek}, where the corresponding entropy was calculated. Other recent works have studied \cite{Horava} BH in the context of Ho\u{r}ava-Lifshitz gravity as well.
Previous works concerning BH in $f(R)$ theories proved that for a Lagrangian $R+aR^2$ the only spherically symmetric solution is Schwarzschild's one provided that one works in the Einstein's frame. Again in Einstein's frame, \cite{Mignemi} proposed uniqueness theorems for spherically symmetric solutions with an arbitrary number of dimensions (see \cite{Multamaki} for additional results).
Spherical solution with sources were also studied in \cite{olmo} whereas \cite{Nzioki:2009av} developed a new covariant formalism to treat spherically symmetric spacetimes claiming that Schwarzschild solution is not a unique static spherically symmetric solution. Spherically symmetric $f(R)$-Maxwell and $f(R)$-Yang-Mills BH were studied on \cite{Taeyoon1}, confirming the existence of numerical asymptotic solution for the second ones. Concerning axially symmetric solutions, authors in \cite{Capozziello:2009jg} showed that these solutions can be derived by generalizing Newman and Janis method to $f(R)$ theories. An scalar-tensor approach is used in \cite{Myung} to show that Kerr BH are unstable in a subset of $f(R)$ models because of the superradiant instability. In \cite{Palatini_Noether} the entropy of BH is calculated in the Palatini formalism by using the Noether charge approach.
Anti de Sitter ($AdS$) BH have been studied \cite{Cognola} in $f(R)$ models using the Euclidean action method (see, e.g., \cite{Hawking&Page, Witten}) to determine different thermodynamic quantities.
In \cite{Briscese}, the entropy of Schwarzschild-de Sitter ($SdS$) BH is calculated in vacuum for certain cosmologically viable models, and their stability discussed. In \cite{DOMBRIZ} it was proved, in an arbitrary number of dimensions, that the only static spherically symmetric solution --up to second order in perturbations--  for a massive BH in $f(R)$ theories was that of  Schwarzschild-$(A)dS$. In that same investigation, a thermodynamic analysis of Schwarzschild-$(A)dS$ BH was performed for various $f(R)$ models, and it was shown the relation between cosmological and thermodynamic viability.


This work is organized in the following way: first, some general results of $f(R)$ theories in the metric formalism are shown in Section II together with
the widely accepted cosmological viability conditions of $f(R)$ theories. The third section is devoted to the study of the axisymmetric, stationary vacuum solution that describes a massive BH with electric charge and angular momentum in these theories.
In Section IV we study the thermodynamical properties of the obtained solutions, whilst fifth section
analyzes graphically the results of the two previous sections for certain $f(R)$ models.
Finally, we present the conclusions obtained from this work in Section VI.

\section{General Results}

In order to study the possible solutions obtained from any $f(R)$ theory, we start from the action:
\begin{eqnarray}
S=S_g+S_m\,,
\label{S}
\end{eqnarray}
where $S_g$ is the gravitational action:
\begin{eqnarray}
S_g=\frac{1}{16 \pi G}\int \text{d}^{4}x\sqrt{\mid g\mid}\,(R+f(R))\,,
\end{eqnarray}
with $G\equiv\,M_p^{-2}$ Newton's constant (where $M_p$ is Planck's mass), $g$ is the determinant of the metric $g_{\mu\nu}$ ($\mu,\nu\,=\,0,1,2,3$), $R$ is the scalar curvature of the spacetime and $f(R)$ is the function that defines the considered theory. From the matter action term $S_m$, we define the energy momentum tensor as:
\begin{eqnarray}
T^{\mu\nu}=-\frac{2}{\sqrt{\mid g\mid}}\frac{\delta S_m}{\delta
g_{\mu\nu}}\,.
\end{eqnarray}
By performing variations of \eqref{S} with respect to the metric tensor, we obtain that the field equations in metric formalism are:
\begin{eqnarray}
 R_{\mu\nu}(1+f'(R)) - \frac{1}{2}\,g_{\mu\nu}\,(R+f(R))&&\nonumber\\[0.15cm]
+(\nabla_\mu \nabla_\nu-g_{\mu\nu}\Box)f'(R)+8\pi G \,T_{\mu\nu}&=&0\,,
\label{ec_campo}
\end{eqnarray}
with $R_{\mu\nu}$ the Ricci tensor, $\Box\,=\nabla_\beta\,\nabla^\beta$ (where $\nabla$ is the covariant derivative) and $f'(R)=\text{d}f(R)/ \text{d}R$. Taking the trace of this equation yields:
\begin{eqnarray}
 R\,(1+f'(R))-2\,(R+f(R))-3\,\Box\,f'(R)+8\pi G\, T\,=0\,,\nonumber\\
\end{eqnarray}
where $T=T^\mu_{\,\,\,\mu}$. It is interesting to stress that, unlike in GR, vacuum solutions ($T=0$) do not necessarily imply a null curvature $R=0$. From equation \eqref{ec_campo} we obtain the condition for vacuum constant scalar curvature $R=R_0$ solutions:
\begin{eqnarray}
 R_{\mu\nu}\,(1+f'(R_0))-\frac{1}{2}\,g_{\mu\nu}\,(R_0+f(R_0))=0\,.
\end{eqnarray}
and the Ricci tensor becomes proportional to the metric:
\begin{eqnarray}
 R_{\mu\nu}=\frac{R_{0}+f(R_0)}{2(1+f'(R_0))}\,g_{\mu\nu}\,,
\end{eqnarray}
with $1+f'(R_0)\neq\,0$. On the other hand, taking the trace on previous equation we obtain:
\begin{eqnarray}
 R_{0}\,(1+f'(R_0))-2\,(R_{0}+f(R_{0}))\,=\,0\,,
\label{ec_escalar}
\end{eqnarray}
and therefore
\begin{eqnarray}
R_0=\frac{2f(R_0)}{f'(R_0)-1}\,.
\label{ec_curvatura_constante}
\end{eqnarray}

%

\subsection{Viability conditions of $f(R)$ theories}
The basic conditions and restrictions \cite{Pogosian} that are usually imposed to $f(R)$ theories to provide consistent both gravitational and cosmological evolutions are:

\begin{enumerate}
\item $f''(R)\geq 0$ for $R\gg f''(R)$. This is the stability requirement for a high curvature  classical regime \cite{Faraoni} and that of the existence of a matter dominated era in cosmological evolution. A simple physical interpretation can be given to this condition: if an effective gravitational constant $G_{eff}\equiv G/(1+f'(R))$ is defined, then the sign of its variation with respect to $R$, $\text{d}G_{eff}/\text{d}R$, is uniquely determined by the sign of $f''(R)$, so in case $f''(R)<0$, $G_{eff}$ would grow as $R$ does, because $R$ generates more and more curvature itself. This mechanism would destabilize the theory, as it wouldn't have a fundamental state because any small curvature would grow to infinite. Instead, if $f''(R)\geq 0$, a counter reaction mechanism operates to compensate this $R$ growth and stabilize the system.

\item $1+f'(R)>0$. 
This conditions ensures that the effective gravitational constant is positive, as it can be checked from the previous definition of $G_{eff}$.
It can also be seen from a quantum point of view as the condition that avoids the graviton from becoming a ghost \cite{Nunez}.

\item $f'(R)<0$. Keeping in mind the strong restrictions of Big Bang nucleosynthesis and cosmic microwave background, this condition ensures GR behavior to be recovered at early times, that is, $f(R)/R\rightarrow0$ and $f'(R)\rightarrow0$  as $R\rightarrow\infty$. Conditions 1 and 2 together demand $f(R)$ to be a monotone increasing function between the values $-1<f'(R)<0$.

\item $f'(R)$ must be small in recent epochs. This condition is mandatory in order to satisfy imposed restrictions by local (solar and galactic) gravity tests. As the analysis done in \cite{Sawicki} indicates, the value of $|f'(R)|$ must not be bigger than $10^{-6}$ (although there is still some controversy about this). This is not a needed requirement if the only goal is to obtain a model that explains cosmic acceleration.
\end{enumerate}


\section{Kerr-Newman Black Holes in $f(R)$ Theories}
Since we are looking for constant curvature $R_0$
vacuum solutions for fields generated by massive charged
objects, the appropriate action (in 
$G=c=\hbar=k_B=1$ units)  is:

\begin{eqnarray}
S&=&\frac{1}{16 \pi}\int \text{d}^{4}x\sqrt{\mid g\mid}\,(R+f(R)-F_{\mu\nu}F^{\mu\nu})\,,
\end{eqnarray}
where $F_{\mu\nu}=\partial_\mu\,A_\nu-\partial_{\nu}A_\mu$ and $A_\mu$ the electromagnetic potential. This action leads to the field equations:
\begin{eqnarray}
 R_{\mu\nu}\,(1+f'(R_0))-\frac{1}{2}\,g_{\mu\nu}\,(R_0+f(R_0))&&\nonumber\\[0.15cm]
-2\left( F_{\mu\alpha}F_{\,\,\,\nu}^\alpha-\frac{1}{4}g_{\mu\nu}F_{\alpha\beta}F^{\alpha\beta}\right) \,&=&\,0\,.
\label{ec_tensorial}
\end{eqnarray}
At this stage, it is worth stressing that if we take the trace of the previous equation, \eqref{ec_escalar} is recovered due to the fact that $F^{\mu}_{\,\,\mu}=0$.

The axisymmetric, stationary and constant curvature $R_0$ solution that describes a BH with mass, electric charge and angular momentum was found by Carter and published for the first time in 1973 \cite{Carter}. In Boyer-Lindquist coordinates, the metric
describing with no coordinate singularities the spacetime exterior
to the BH and interior to the cosmological horizon (provided it exists, as will be studied below), takes the form:
\begin{widetext}
\begin{eqnarray}
\text{d}s^2\,=\,\frac{\rho^2}{\Delta_r}\,\text{d}r^2+\frac{\rho^2}{\Delta_\theta}\,\text{d}\theta^2+\frac{\Delta_\theta\,\sin^2{\theta}}{\rho^2}\,\left[a\,\frac{\text{d}t}{\Xi}-\left(r^2+a^2\right)\,\frac{\text{d}\phi}{\Xi}\right]^2-\frac{\Delta_r}{\rho^2}\,\left(\frac{\text{d}t}{\Xi}-a\sin^2{\theta}\frac{\text{d}\phi}{\Xi}\right)^2\,,
\label{metrica}
\end{eqnarray}
\end{widetext}
with:
\begin{eqnarray}
\Delta_r\,&:=&\,\left(r^2+a^2\right)\left(1-\frac{R_0}{12}\,r^2\right)-2Mr+\frac{q^2}{\left( 1+f'(R_0 )\right)}\,,\nonumber
\end{eqnarray}
\begin{eqnarray}
\rho^2\,&:=&\,r^2+a^2\cos^2{\theta}\,,\nonumber
\end{eqnarray}
\begin{eqnarray}
\Delta_\theta\,&:=&\,1+\frac{R_0}{12}\,a^2\cos^2{\theta}\,,\nonumber
\end{eqnarray}
\begin{eqnarray}
\Xi\,&:=&\,1+\frac{R_0}{12}\,a^2\,,
\label{definiciones}
\end{eqnarray}
where $M$, $a$ and $q$ denote the mass, spin and electric charge parameters respectively. Notice that, unlike in the GR case, the contribution of the charge of the BH to the metric is corrected by a $\left(1+f'(R_0) \right)^{-1/2}$ factor. This feature was already obtained for Reissner–Nordström BH in \cite{DOMBRIZ}.

On the other hand, the required potential vector and electromagnetic field tensor in equation \eqref{ec_tensorial} solutions
for metric \eqref{metrica} are respectively:
\begin{eqnarray}
A&=&-\frac{q\,r}{\rho^2}\left(\frac{\text{d}t}{\Xi}-a\sin^2{\theta}\frac{\text{d}\phi}{\Xi}\right),\nonumber\\[0.15cm]
F&=&-\frac{q\,(r^2-a^2\cos^2{\theta})}{\rho^4}\left(\frac{\text{d}t}{\Xi}-a\sin^2{\theta}\frac{\text{d}\phi}{\Xi}\right)\wedge\text{d}r\nonumber\\[0.15cm]
&&-\frac{2\,q\,r\,a\cos{\theta}\sin{\theta}}{\rho^4}\,\text{d}\theta\wedge\left[a\,\frac{\text{d}t}{\Xi}-(r^2+a^2)\,\frac{\text{d}\,\phi}{\Xi}\right]\,.\nonumber\\[0.2cm]
\label{EM}
\end{eqnarray}
To lighten notation, from now on we will use $Q^2\equiv q^2\, /\,\left(1+f'(R_0) \right)$ to refer to the electric charge parameter of the BH.

The nature of coordinates in \eqref{metrica} can be seen by considering the $M\rightarrow0$, $Q\rightarrow 0$, $R_0\rightarrow 0$ limits on this
metric. Thus \eqref{metrica} becomes:
\begin{eqnarray}
\text{d}s^2=-\text{d}t^2+\frac{\rho^2}{r^2+a^2}\,\text{d}r^2+\rho^2\text{d}\theta^2+(r^2+a^2)\sin^2{\theta}\,\text{d}\phi^2\,,\nonumber\\
\label{Boyer_Lindquist_Minkowski}
\end{eqnarray}

i.e., Minkowski spacetime in spacial coordinates $(r,\,\theta,\,\phi)$. It is not obvious from \eqref{Boyer_Lindquist_Minkowski} that what one has is Minkowski spacetime, this is because Boyer-Lindquist coordinates need to be ``untwisted'' via cartesian coordinates to confirm that what we have is actually an empty spacetime:
\begin{eqnarray}
x&=&\sqrt{r^2+a^2}\,\sin{\theta}\,\cos{\phi}\,,\nonumber\\[0.1cm]
y&=&\sqrt{r^2+a^2}\,\sin{\theta}\,\sin{\phi}\,,\nonumber\\[0.1cm]
z&=&r\,\cos{\theta}\,,
\end{eqnarray}
with $r\geq0$, $0\leq\theta\leq\pi$ y $0\leq\phi\leq2\pi$. Nevertheless, one must keep in mind that, when $M\neq 0,\,Q\neq 0$ and $R_0\neq 0$, the simplest
interpretation given to these coordinates is not completely appropriate due to the distortion of the empty spacetime that the presence of the BH induces. On the other hand, if we do $M\rightarrow0$, $Q\rightarrow 0$, $a\rightarrow 0$, we obtain a constant curvature spacetime metric:
\begin{eqnarray}
\text{d}s^2=-\left(1-\frac{\displaystyle R_0\,r^2}{\displaystyle 12}\right)\text{d}t^2+\displaystyle \frac{\displaystyle1}{\left(\displaystyle1-\displaystyle \frac{R_0\,r^2}{12}\right)}\,\text{d}r^2+r^2\text{d}\Omega^{2}_{(2)}\,,\nonumber\\
\end{eqnarray}

that corresponds to either $dS$ or $AdS$ spacetime depending on the sign of $R_0$. It is also easy to verify that when
$a\rightarrow0$, $Q\rightarrow0$, Schwarzschild-$(A)dS$ BH is recovered.

\subsection{Singularities}

We will study now the singularities of these BH. Calculating $R^{\mu\nu\sigma\rho}\,R_{\mu\nu\sigma\rho}$, only $\rho=0$ happens to be an intrinsic singularity, and considering the definition of $\rho$ in \eqref{definiciones}, such singularity is given by:
\begin{eqnarray}
r=0\,\,\,\,\,\,\,\,\,{\text {and}}\,\,\,\,\,\,\,\,\theta=\pi/2\,.
\end{eqnarray}\\[-0.2cm]
Keeping in mind that we are working with Boyer-Lindquist coordinates, the set of points given by $r=0$ and $\theta=\pi\,/2$  represent a ring in the equatorial plane of radius $a$ centered on the rotation axis of the BH, just as it happens in Kerr BH \cite{kerr}.
\begin{figure*}[htp]
	\centering
		\includegraphics[width=0.38\textwidth]{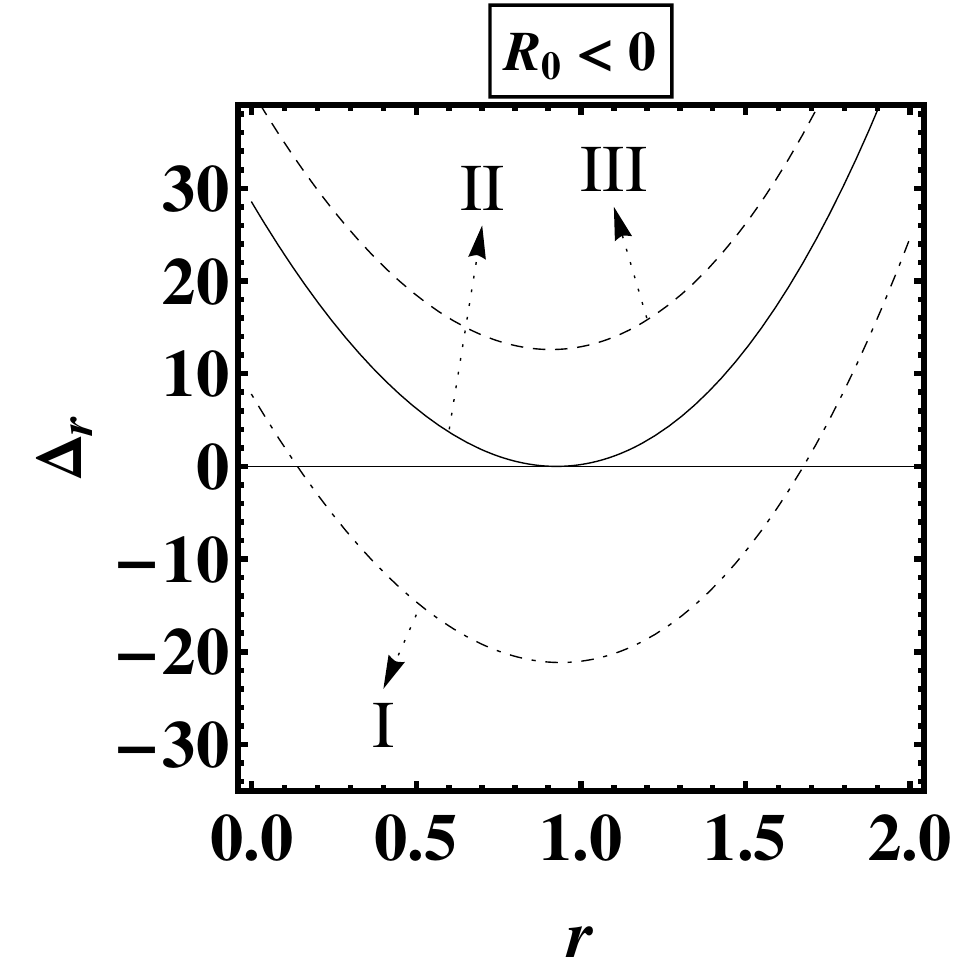}\,\,\,\,\,\,\,\,\,\,\,\,\,\,
		\includegraphics[width=0.38\textwidth]{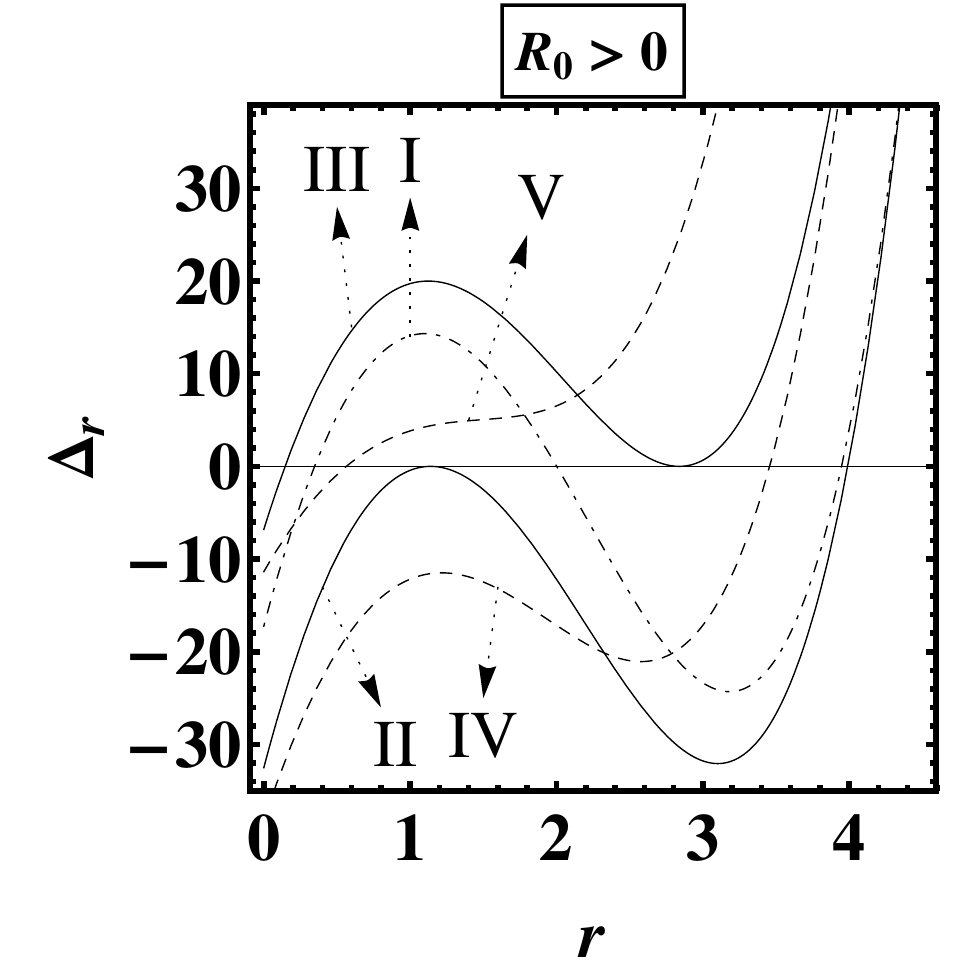}
		\caption{\footnotesize{
		Graphics showing positions of horizons as solutions of the equation $\Delta_r=0$. On the left panel ($R_0<0$) we show the cases $h>0$ ({\bf I}, BH with well-defined horizons, dashed with dots), $h=0$ ({\bf II}, {\it extremal} BH, continuous line) and $h<0$ ({\bf III}, {\it naked singularity}, dashed). On the right panel ($R_0>0$) we represent the cases $h<0$ ({\bf I}, BH with well-defined horizons, dashed with dots), $h=0$ ({\bf II}, $extremal$ BH and {\bf III}, $extremal$ $marginal$ BH, continuous line), and $h>0$ ({\bf IV}, {\it naked singularity} and {\bf V}, {\it naked marginal singularity}, dashed).}}
	\label{fig:graficashor}
\end{figure*}

\subsection{Horizons}

It is also interesting to study the horizon structure of these BH: according to the {\it horizon} definition $g^{rr}=0$, they are found as the roots
of the equation $\Delta_r=0$, that is:
\begin{eqnarray}
r^4+\left(a^2-\frac{12}{R_0}\right)\,r^2+\frac{24\,M}{R_0}\,r-\frac{12}{R_0}\,\left(a^2+Q^2\right)\,=\,0\,,\nonumber\\[-0.2cm]
\label{poli_hori}
\end{eqnarray}
fourth order equation that can be rewritten as:
\begin{eqnarray}
(r-r_{-})(r-r_{int})(r-r_{ext})(r-r_{cosm})\,=\,0\,,
\end{eqnarray}
where $r_-$ is always a negative solution with no physical meaning, $r_{int}$ and $r_{ext}$ are the interior and exterior horizon respectively, and $r_{cosm}$ represents -- provided it arises, as will be  seen later -- the cosmological event horizon for observers between $r_{ext}$ and $r_{cosm}$. This horizon divides the region that the observer could see from the region he could never see if he waited long enough time. Using L. Ferrari's method \cite{Ludovico} to solve quartic equations, the existence 
of real solutions for this equation is given by a factor $h$ let us name it {\it horizon parameter}:
\begin{widetext}
\begin{eqnarray}
h\equiv\left[\frac{4}{R_0}\left(1-\frac{R_0}{12}\,a^2\right)^2-4\,\left(a^2+Q^2\right)\right]^3+\frac{4}{R_0}\left\{\left(1-\frac{R_0}{12}\,a^2\right)\left[\frac{4}{R_0}\left(1-\frac{R_0}{12}\,a^2\right)^2+12\,\left(a^2+Q^2\right)\right]-18\,M^2\right\}^2\,.\nonumber\\
\label{ecu_D}
\end{eqnarray}
\end{widetext}

For a negative scalar curvature $R_0$, three options may be considered: $i)$ $h>0$: there are only two real solutions, $r_{int}$ and $r_{ext}$, lacking this configuration a cosmological horizon, as it is expected for an $AdS$ like Universe. $ii)$ $h=0$: there is only a degenerated root, particular case of an {\it extremal} BH, whose interior and exterior horizons have merged into one single horizon with a null surface gravity $\kappa$ (that will be defined in section IV). $iii)$ $h<0$: it is found that there is no real solution to \eqref{ecu_D}, which translates into an absence of horizons that leads to a {\it naked singularity}.

For a positive curvature $R_0$, there are also several configurations depending on the value of $h$:
$i)$ $h<0$: both $r_{int}$, $r_{ext}$ and $r_{cosm}$ are positive and real, thus the BH possesses a well-defined horizon structure in an Universe with a cosmological horizon.
$ii)$ $h=0$: two different cases may be described, either $r_{int}$ and $r_{ext}$ become degenerated solutions, or $r_{ext}$ and $r_{cosm}$ do so. The first case represents an $extremal$ BH, described before. The second one can be understood as the cosmological limit for which a BH preserves its exterior horizon without being "torn apart" due to the relative recession speed between two radially separated points induced by the cosmic expansion in an Universe described by a constant positive curvature; this case is known as {\it marginal naked singularity}.
$iii)$ $h>0$: There is only one positive root, that may be either $r_{int}$ or $r_{cosm}$. In the first case, the mass of the BH has exceeded the limit imposed by the cosmology (just described for $h=0$), and there are neither exterior nor cosmological horizon. This situation just leaves the interior horizon to cover the singularity ({\it marginal naked singularity case}. On the contrary, if the root corresponds to $r_{cosm}$, this time there is a {\it naked singularity} with a cosmological horizon.
In Figure \ref{fig:graficashor} we have shown the zeros of the fourth order polynomial $\Delta_r$:
 \begin{figure*}[htp]
	\centering
		\includegraphics[width=0.38\textwidth]{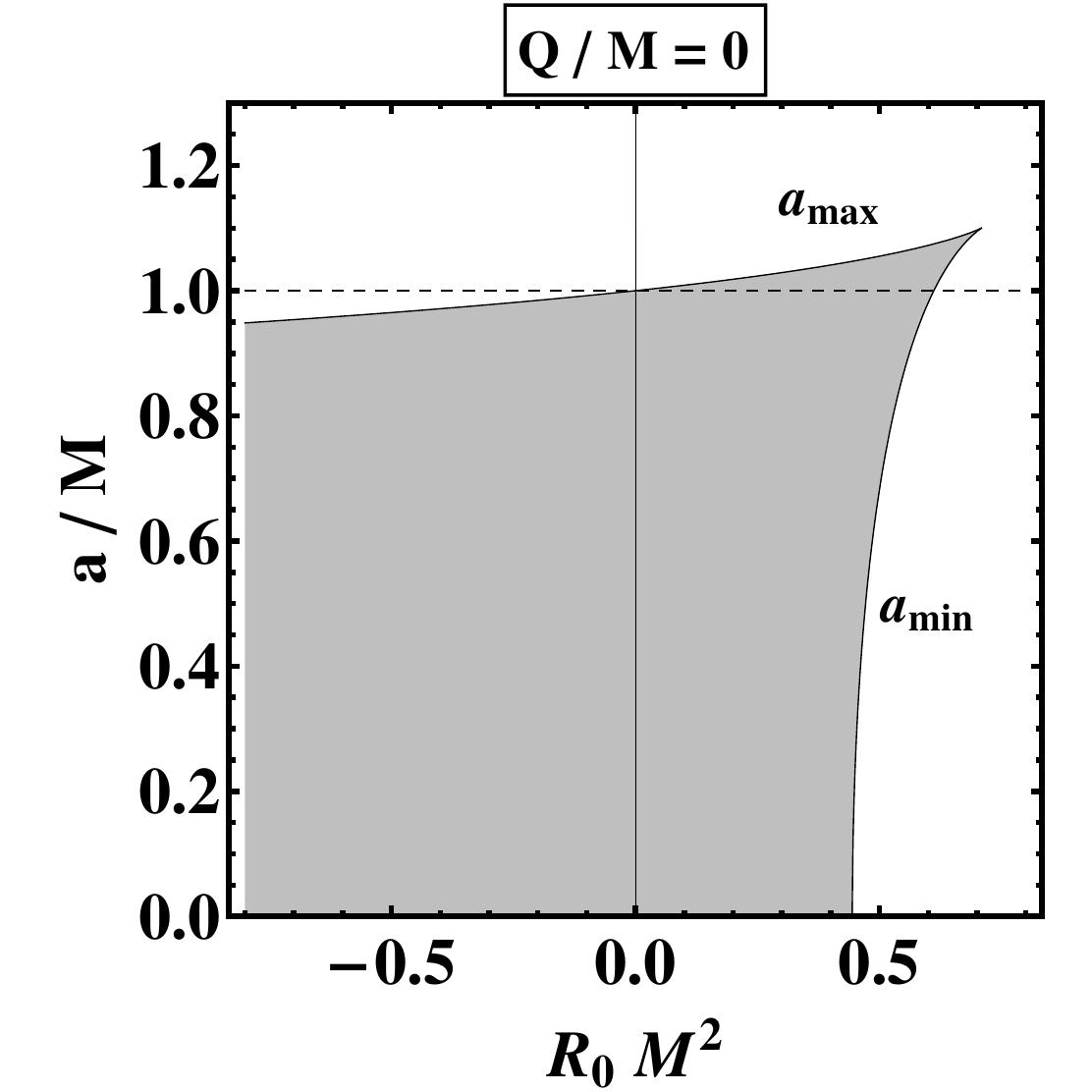}\,\,\,\,\,\,\,\,\,\,\,\,\,\,
		\includegraphics[width=0.38\textwidth]{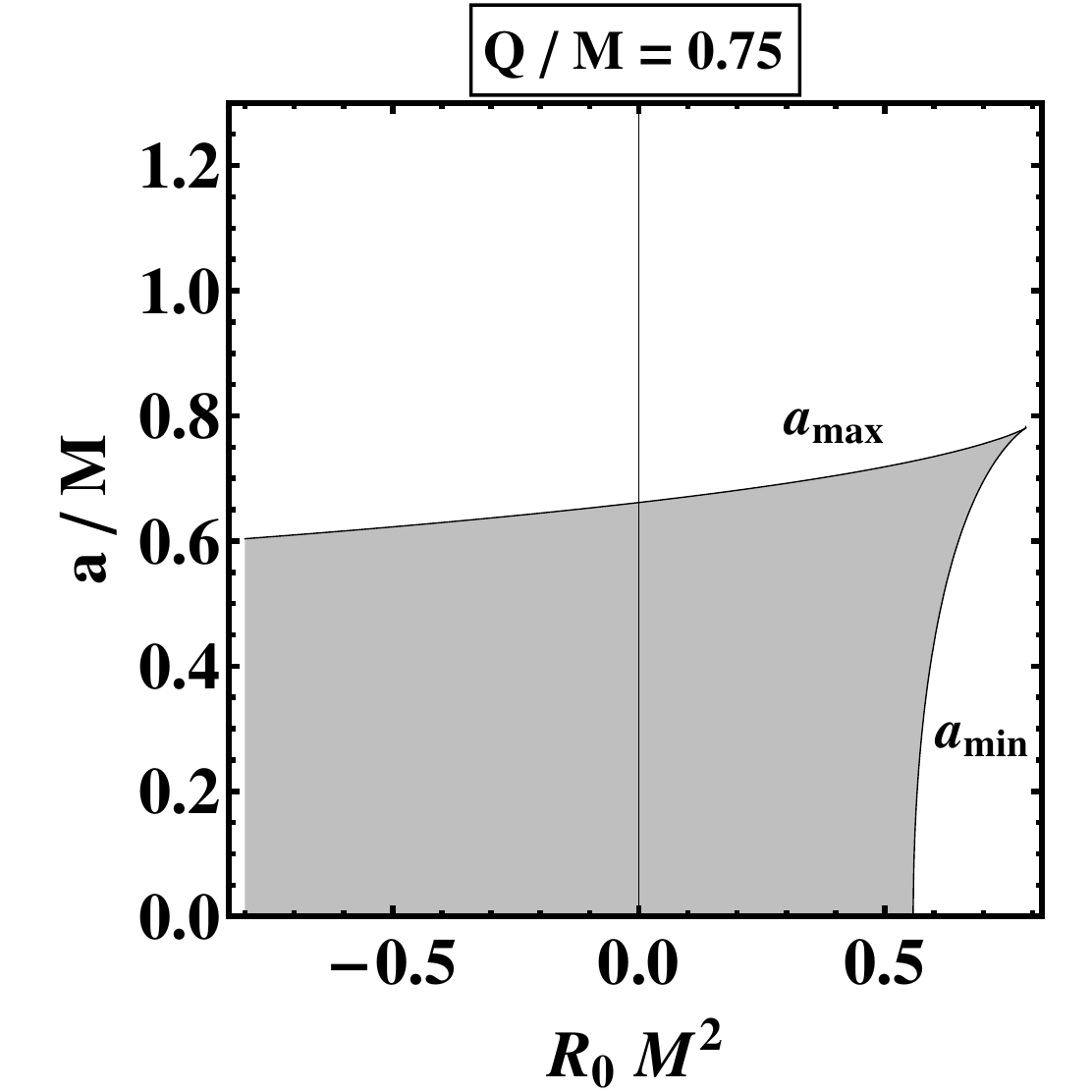}
		\caption{\footnotesize{The shaded regions, delimited by the upper $a_{max}$ and lower $a_{min}$ curves, represent the values of $a/M$
		for which the existence of BH is possible once $R_0\,M^2$ value is fixed. It is shown for $Q\,/\,M=0$ and $Q\,/\,M=0.75$ on the left and right panels repectively. Note that $R_0$ has dimensions of [length]$^{-2}$ when normalizing.}}
	\label{fig:rangoespin}
\end{figure*}
it is remarkable that, from a certain positive value of the curvature $R_{0}^{\,crit}$ onward, the $h$ factor goes to zero for two values of $a$, i.e., apart from the usual $a_{max}$ for which the BH turns $extremal$, there is now a spin lower bound $a_{min}$, below which the BH turns into a {\it marginal extremal} BH, as we discussed before. Therefore
\begin{eqnarray}
h\,(a_{max},\,M,\,|R_0|\geq0,\,Q)=0\,\,\,\,\,\,\nonumber\\[0.15cm]
\Rightarrow\,\, a_{max} \equiv a_{max} \, (M,\,|R_0|\geq 0,\,Q)\,,
\end{eqnarray}
\begin{eqnarray}
h\,(a_{min},\,M,\,R_0\geq R_{0}^{\,crit}>0,\,Q)=0\,\,\,\,\,\,\nonumber\\[0.15cm]
\Rightarrow \,\,a_{min} \equiv a_{min} \, (M,\,R_0\geq R_{0}^{\,crit}>0,\,Q)\,.
\end{eqnarray}
Due to the excessive length
of the equations that describe
the behavior of $a_{max}$ and $a_{min}$, we prefer not
to display them here. Instead, in Figure \ref{fig:rangoespin} we show --for certain values of the electric charge $Q$ parameter $Q$ --
the range of values of the spin $a$ parameter for which
BH are allowed taking into account. To do so, the corresponding $R_0$ value
is determined by the parameters defining each $f(R)$ model, as can be
seen from equation \eqref{ec_escalar}.

\subsection{Stationary Limit Surfaces}

\begin{figure*}[tp]
	\centering
		\includegraphics[width=0.4\textwidth]{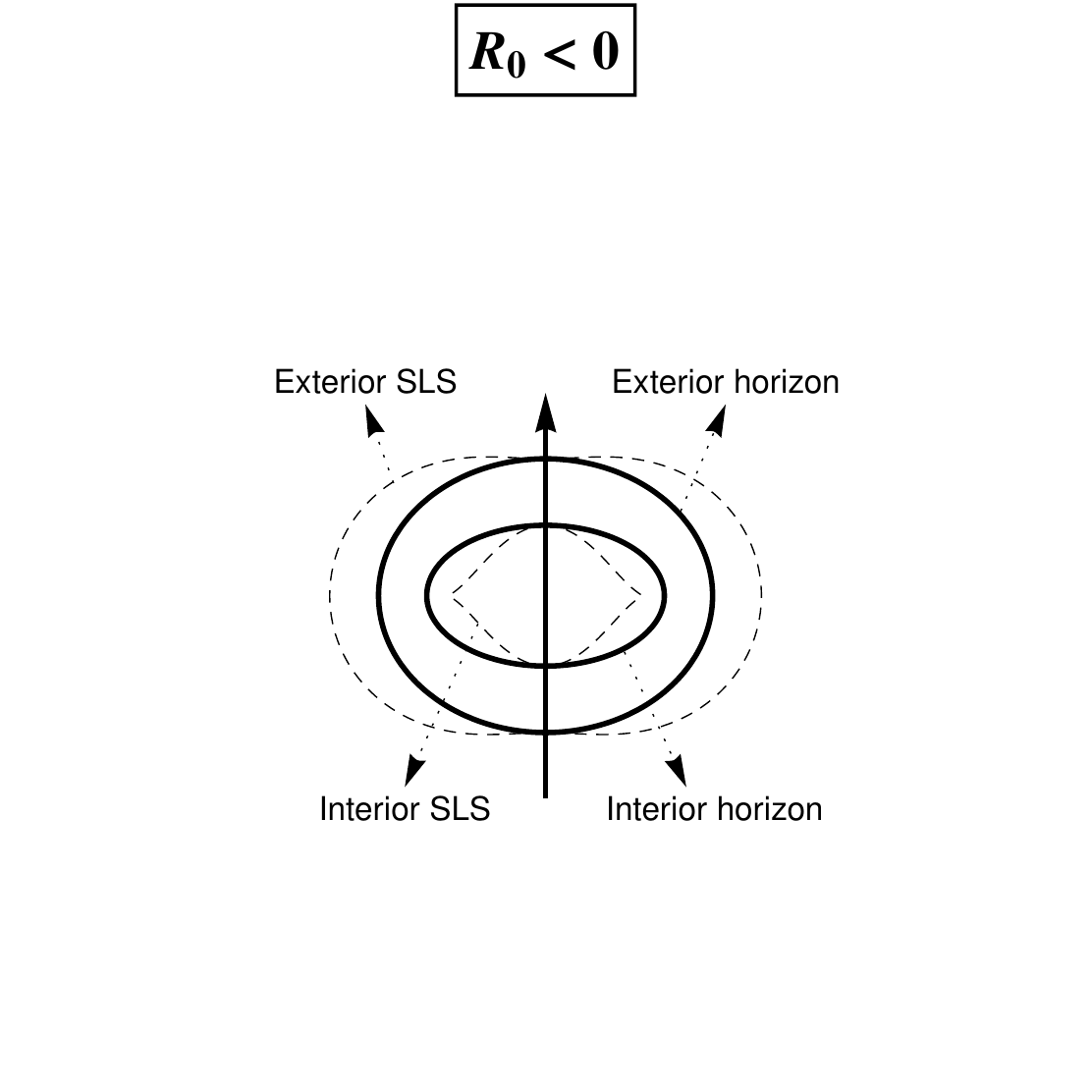}
		\includegraphics[width=0.4\textwidth]{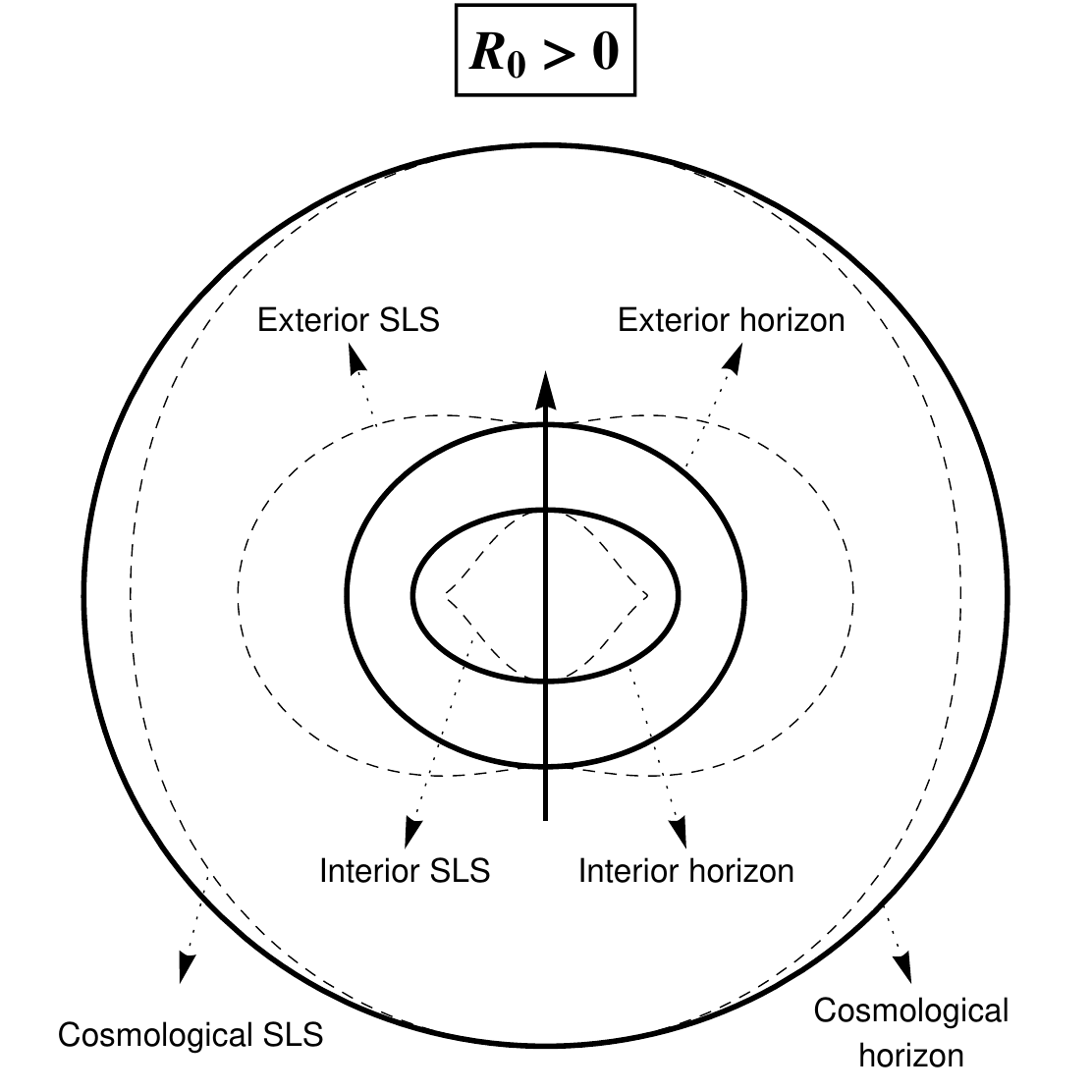}
		\caption{\footnotesize{
On the left: diagram of a Kerr-Newman BH structure with negative curvature solution $R_0=-0.4<0$, $M=1$, $a=0.85$ and $Q=0.35$ ($h>0$). Right: BH structure with positive curvature solution $R_0=0.4>0$, $M=1$, $a=0.9$ and $Q=0.4$ ($h<0$). Dotted surfaces represent the SLS whereas horizons are shown with continuous lines. The rotation axis of the BH is indicated by the vertical arrow. In both types of BH, the region between the exterior SLS $r_{S\,ext}$  and its associated exterior horizon $r_{ext}$ is known as {\it ergoregion}.}}
	\label{fig:agujerokerrpng}
\end{figure*}

Another interesting feature of Kerr-Newman BH are Stationary Limit Surfaces (SLS), given by $g_{tt}=0$. For Boyer-Lindquist coordinates, this condition translates into:
\begin{eqnarray}
\frac{\Delta_\theta\,\sin^2{\theta}\,a^2}{\rho^2\,\Xi^2}-\frac{\Delta_r}{\rho^2\,\Xi^2}=0,
\end{eqnarray}
that leads to the fourth order equation
\begin{widetext}
\begin{eqnarray}
r^4+\left(a^2-\frac{12}{R_0}\right)\,r^2+\frac{24\,M}{R_0}\,r-\left(a^2\,\cos^2{\theta}+\frac{12}{R_0}\right)\,a^2\,\sin^2{\theta}-\frac{12}{R_0}\,(a^2+Q^2)=\,0,
\label{ecu_sup}
\end{eqnarray}
\end{widetext}
which can be rewritten as:
\begin{eqnarray}
(r-r_{S\,-})(r-r_{S\,int})(r-r_{S\,ext})(r-r_{S\,cosm})=0.
\end{eqnarray}
From this equation it follows that each horizon has one "associated" SLS. Both hypersurfaces  coincide at $\theta=0,\pi$  as seen when comparing \eqref{ecu_sup}
with equation (\ref{poli_hori}). A complete
scheme of BH horizons and
SLS structure is shown in Figure \ref{fig:agujerokerrpng} for both signs of $R_0$.

%
\section{Black Hole Thermodynamics}

From now on, we will focus our study on BH with a well-defined horizon structure and only for $R_0$ negative values of. This last choice is motivated
by the problems arising when normalizing the temporal Killing $\xi\equiv\partial_t$ in positive curvature spacetimes. This problem is more extensively examined in \cite{Cosmohorizon}. $R_0<0$ choice will allow us to
define the thermodynamical quantities of the BH. The rotation Killing field $\psi\equiv\partial_{\phi}$ is uniquely determined by the condition that their orbits should be closed curves with a length parameter equal to $2\pi$. Nevertheless, there is not an adequate criterion to normalize the Killing vector $\xi\equiv\partial_t$ in the $dS$ ($R_0>0$) Universe since multiplicative constants can be added to $\xi$ and the obtained Killing vectors continue being null on the horizon. In the
$AdS$ ($R_0<0$), the normalization is done without problems by imposing that the $\xi$ value tends to
$\,r\left(-12\,/\,R_0\right)^{1/2}$ as $r$ goes to infinity.

In order to study the different thermodynamical properties of Kerr-Newman BH in $f(R)$ theories, we start looking for the temperature of the exterior horizon $r_{ext}\equiv r_{ext}\left(R_0,\,a,\,Q,\,M\right)$.  For that purpose, we will use the Euclidean action method \cite{HGG}. Performing the change $t\rightarrow -i\tau$, $a\rightarrow ia$ on the metric \eqref{metrica} we obtain the Euclidean section, whose non singular metric is positive-definite, and time coordinate has now angular character around the ``axis'' $r=r_{ext}$. Regularity of the metric at $r=r_{ext}$ requires the identification of points:
\begin{eqnarray}
(\tau,\,r,\,\theta,\,\phi)\, \sim \,(\tau+i\,\beta,\,r,\,\theta,\,\phi+i\,\beta\,\Omega_H),
\end{eqnarray}
where $\beta$, which represents the period of the imaginary time on the Euclidean section, it's the inverse Hawking temperature:
\begin{eqnarray}
\beta= \frac{\displaystyle 4\pi\left(r_{ext}^2+a^2\right)}{\displaystyle r_{ext}\left[1-\frac{R_0\,a^2}{12}-\frac{R_0\,r_{ext}^2}{4}-\frac{\left(a^2+Q^2\right)}{r_{ext}^2}\right]}\equiv\frac{1}{T_E},\nonumber\\
\label{inversatemp}
\end{eqnarray}
and $\Omega_H$ is the angular velocity of the rotating horizon, which is the same for all the horizon:
\begin{eqnarray}
\Omega_H=\frac {a\,\Xi}{\displaystyle r_{ext}^2+a^2}.
\label{velangular}
\end{eqnarray}
Considering that the event horizon is also a Killing horizon of the Killing vector $\chi=\xi+\Omega_H\psi$ (where, as was said before, $\xi\equiv\partial_t$ and $\psi\equiv\partial_{\phi}$ are the vectors that asymptotically represent time translations and rotations respectively), $\Omega_H$ could be also obtained demanding $\chi$ to be a null vector on the horizon:
\begin{eqnarray}
\left.\chi^\mu\chi_\mu\right|_{r=r_{ext}}=0
\end{eqnarray}

BH horizon temperature could have also been obtained through Killing vectors, as is explained in \cite{Hawking1974} where temperature is defined as follows:
\begin{eqnarray}
T_{\kappa}\equiv\frac{\kappa}{4\,\pi},
\end{eqnarray}
with $\kappa$ the surface gravity defined by:
\begin{eqnarray}
\chi^\mu\,\nabla_\mu\chi_\nu=\kappa\,\chi_\nu.
\end{eqnarray}
It can be verified that $\kappa$ is the same at any horizon point and consequently $T_\kappa=T_E$ as obtained in \cite{Bardeen1973}.

Now that we know the expression for the temperature, we consider the Euclidean action in order to obtain the remaining thermodynamical quantities:
\begin{eqnarray}
\Delta S_E=\frac{1}{16 \pi}\int _{\cal Y}\text{d}^{4}x\sqrt{\mid g\mid}\,\left(R_0+f(R_0)-F_{\mu\nu}F^{\mu\nu}\right)\,,\nonumber\\
\end{eqnarray}
with ${\cal Y}$ the integration region. As is described in \cite{Hawking&Page}, one has to calculate the difference in four-volumes of the two metrics, identified by the same imaginary time. Provided that the metric is stationary, integration over time simply leads to a multiplicative factor $\beta$. On the other hand, keeping in mind that Maxwell's
equations $\nabla_{\nu} F^{\mu\nu}=0$ must be satisfied, we can rewrite the third term in the integrand as a divergence:
\begin{eqnarray}
 F_{\mu\nu}F^{\mu\nu}=\nabla_{\nu}(2\,F^{\mu\nu}\,A_{\mu}),
\end{eqnarray}
and therefore:
\begin{eqnarray}
\Delta S_E=\frac{R_0+f(R_0)}{16\,\pi}\,\beta\,\Delta \mbox{V}+\frac{1}{8 \pi}\int  _{\cal{\partial Y}}F^{\mu\nu}\,A_{\mu}\,\text{d}\Sigma_{\nu}\,,\nonumber\\
\end{eqnarray}
where $\partial \cal{Y}=$ $S^1 \times S^2$ is the boundary of the considered region, $S^2$ is a 2-sphere whose radius has to be sent to infinity after the integration, and $\Delta \mbox{V}$ is the volume difference between both solutions (corresponding to the black hole metric and that of $AdS$ space identified with the same imaginary time). After some calculation, we obtain:
\begin{widetext}
\begin{eqnarray}
\Delta S_E=\frac{ \beta\,\left(R_0+f(R_0)\right)}{24\,\Xi}
\left[
r_{ext}^3+\left(a^2+\frac{12}{R_0}\right)\,r_{ext}+\frac{12\,a^2}{R_0\,r_{ext}}
\right]
+
\frac{\beta}{2}\,\Phi_e\,{\cal{Q}}\,\left(\frac{r_{ext}^2+a^2}{2\,r_{ext}^2}+1\right),
\label{accion}
\end{eqnarray}
\end{widetext}
where $\Phi_e$ is the electric potential of the horizon as seen from infinity:
\begin{eqnarray}
\Phi_e=\left.A_\mu\chi^\mu\right|_{r\rightarrow \infty}-\left.A_{\mu}\chi^\mu\right|_{r=r_{ext}}=\frac{q\,r_{ext}}{r_{ext}^2+a^2}\,,
\end{eqnarray}
and ${\cal{Q}}$ is the physical electric charge of the BH, obtained integrating the flux of the electromagnetic field tensor at infinity, which happens to be:
\begin{eqnarray}
{\cal{Q}}=\frac{q}{\Xi}\,.
\label{cargafisica}
\end{eqnarray}
We shall remember that these calculations involve the vector potential and the electromagnetic field tensor given in \eqref{EM}, and that's why the factor $\left(1+f'(R_0) \right)^{-1/2}$ does not appear here. Further analysis of the action reveals that it goes singular for $h=0$, as could be expected from {\it extremal} BH, whose temperature $T_E=0$ makes the $\beta$ factor diverge. Since thermodynamical potentials are obtained by dividing the action by the $\beta$ factor, they still remain well defined at $T_E=0$.

It can also be seen that the action \eqref{accion} diverges in the limit $a^2=-12/R_0$, which implies $\Xi=0$. This singular case \cite{Footnote1}
is further explored in \cite{RotationAdS} and implies that a 3-dimensional static closed Universe at
infinity would rotate with the speed of light. Thus, in order to avoid all these problematic issues, let us assume from now on that:
\begin{eqnarray}
\Xi\,:=\,\left(1+\frac{R_0\,a^2}{12}\right)>0.
\label{condicionrotacion}
\end{eqnarray}
The previous expression will turn out also to be a required
condition to ensure a positive area and entropy of the BH, as will be seen below.

By using the expression \eqref{accion} we can immediately obtain Helmholtz free energy $F$, defined by:
\begin{eqnarray}
F=\frac{\Delta S_E}{\beta}+\Omega_H\,J,
\end{eqnarray}
where the term $\Omega_H\,J$ comes from the required Legendre transformation to fix angular momentum, being $J$ the angular momentum of the BH and $\Omega_H$ the angular velocity of the horizon computed before in \eqref{velangular}. To calculate $J$ we need first the physical mass associated to the BH, which can be calculated from:
\begin{eqnarray}
{\cal M}=\frac{\partial \Delta S_E}{\partial \beta}=\frac{M}{\Xi}\,\left(1+f'(R_0)\right)\,,
\end{eqnarray}
resulting on an angular momentum:
\begin{eqnarray}
J=\frac{a\,{\cal M}}{\Xi}=\frac{a\,M}{\Xi^2}\,\left(1+f'(R_0)\right)\,,
\end{eqnarray}
where we have used equation \eqref{ec_escalar} on ${\cal M}$ to make the substitution: $2\,(R_0+f(R_0))\,/\,R_0 = 1+f'(R_0)$. Using again the relation $Q^2=q^2\, /\,\left(1+f'(R_0) \right)$, and equation \eqref{poli_hori} to express $M$ as a function of $r_{ext}$, we obtain:
\begin{widetext}
\begin{eqnarray}
F=(1+f'(R_0))\frac{\displaystyle \left[36\,Q^2+12\,r_{ext}^2+r_{ext}^4\,R_0+a^2\,(36-r_{ext}^2\,R_0)\right]}{\displaystyle  24\,r_{ext}\,\Xi}\,.
\end{eqnarray}
\end{widetext}
If the condition $1+f'(R_0)>0$ is required to hold in order to obtain positive values of the mass, by analyzing the numerator of $F$, we find
$F>0$ for values of the horizon below $r_{ext}^{\,\text{\it limit}}$ (with an associated mass $M^{\,\text{\it limit}}$ through equation \eqref{poli_hori}), and $F<0$ for larger values.

Using the appropriate thermodynamical relations \cite{Caldarelli}, we can derive the entropy $S$ of the BH, which reads:
\begin{eqnarray}
S&=&\beta\,({\cal{M}}-\Omega_H\,J)-\Delta S_E\nonumber\\[0.15cm]
&=&(1+f'(R_0))\,\frac{\pi\,(r_{ext}^2+a^2)}{\displaystyle \Xi}.
\label{entropia}
\end{eqnarray}
If we compute now the area ${\cal{A}}_H$ of the exterior horizon $r_{ext}$, which can be calculated from the metric \eqref{metrica} doing $r=r_{ext}$ and
$t$ constants, we obtain:
\begin{eqnarray}
\text{d}s^2_2\,=\,\frac{\rho^2}{\Delta_\theta}\,\text{d}\theta^2+\frac{\Delta_\theta\,\sin^2{\theta}}{\rho^2}\,\left[-(r_{ext}^2+a^2)\,\frac{\text{d}\,\phi}{\Xi}\right]^2
\end{eqnarray}
\begin{eqnarray}
{\cal{A}}_H\equiv\int\int\sqrt{\text{det\,g}_{(2)}}\,\text{d}\theta\,\text{d}\phi=\frac{4\pi(r_{ext}^2+a^2)}{\Xi}
\end{eqnarray}
Therefore one sees straightforwardly that the entropy \eqref{entropia} can be expressed as:
\begin{eqnarray}
S=(1+f'(R_0))\,\frac{ {\cal A}_H }{4}\,,
\end{eqnarray}
consequently $1+f'(R_0)>0$ is also a mandatory condition to obtain a positive entropy \cite{Footnote2}, as we supposed above.

Once the temperature $T$ and the entropy $S$ of the BH are known, we can take a step further and study
the heat capacity at constant scalar curvature $R_0$ and at fixed spin $a$ and charge $Q$ parameters. From the definition:
\begin{eqnarray}
C=T\left.\frac{\partial S}{\partial T}\right|_{R_0,a,Q},
\end{eqnarray}
we obtain the expression:
\begin{widetext}
\begin{eqnarray}
C=(1+f'(R_0))\,\frac{\displaystyle 2\pi\, r_{ext}^2\,(a^2+ r_{ext}^2)\left[a^2\,(12+ r_{ext}^2\, R_0)+3\,(4\,Q^2-4\, r_{ext}^2+ r_{ext}^4 \,R_0)\right]}{\displaystyle \Xi \left[-36\,Q^2\,r_{ext}^2+a^4\,(-12+r_{ext}^2\, R_0)+3\,r_{ext}^4\,(4+r_{ext}^2\,R_0)-4a^2\,(3\,Q^2+12\,r_{ext}^2-2\,r_{ext}^4\, R_0)\right]}\,.\nonumber\\
\end{eqnarray}
\end{widetext}
Provided that the condition \eqref{condicionrotacion} holds, it seems interesting to find out for which values of $R_0$, $a$, $Q$ and $M$ the denominator of the thermal capacity goes to zero, i.e., the thermal capacity goes through an infinite discontinuity, which corresponds to a BH phase transition. We can distinguish between two kind of BH on this subject depending on the values of the $a$, $Q$ and $M$ parameters and scalar curvature $R_0$: $i)$ {\it fast} BH, without phase transitions and always positive heat capacity $C>0$. $ii)$ {\it slow} BH, presents two phase transitions for two determined values of $r_{ext}$:
\begin{eqnarray}
\left.\frac{\partial T}{\partial r_{ext}}\right|_{R_0,a,Q}(r_{ext}=r_{ext}^{\textnormal I},\,r_{ext}^{\textnormal {II}})\,=\,0,
\end{eqnarray}
with $r_{ext}^{\textnormal {I}}<r_{ext}^{\textnormal {II}}$, being $r_{ext}^{\textnormal I}$ a local maximum of the temperature $T_{max}$, and
$r_{ext}^{\textnormal {II}}$ a local minimum $T_{min}$.
BH heat capacity
proves to be positive ($C>0$) for $r_{ext}<r_{ext}^{\textnormal I}$ and $r_{ext}>r_{ext}^{\textnormal {II}}$, and negative ($C<0$) for $r_{ext}^{\textnormal I}<r_{ext}<r_{ext}^{\textnormal {II}}$. Once again two masses $M^{\textnormal {I}}$ and $M^{\textnormal {II}}$ can be associated  to the radii $r_{ext}^{\textnormal {I}}$ and $r_{ext}^{\textnormal {II}}$ via equation \eqref{poli_hori}.

\begin{figure*}[tp]
	\centering
		\includegraphics[width=0.375\textwidth]{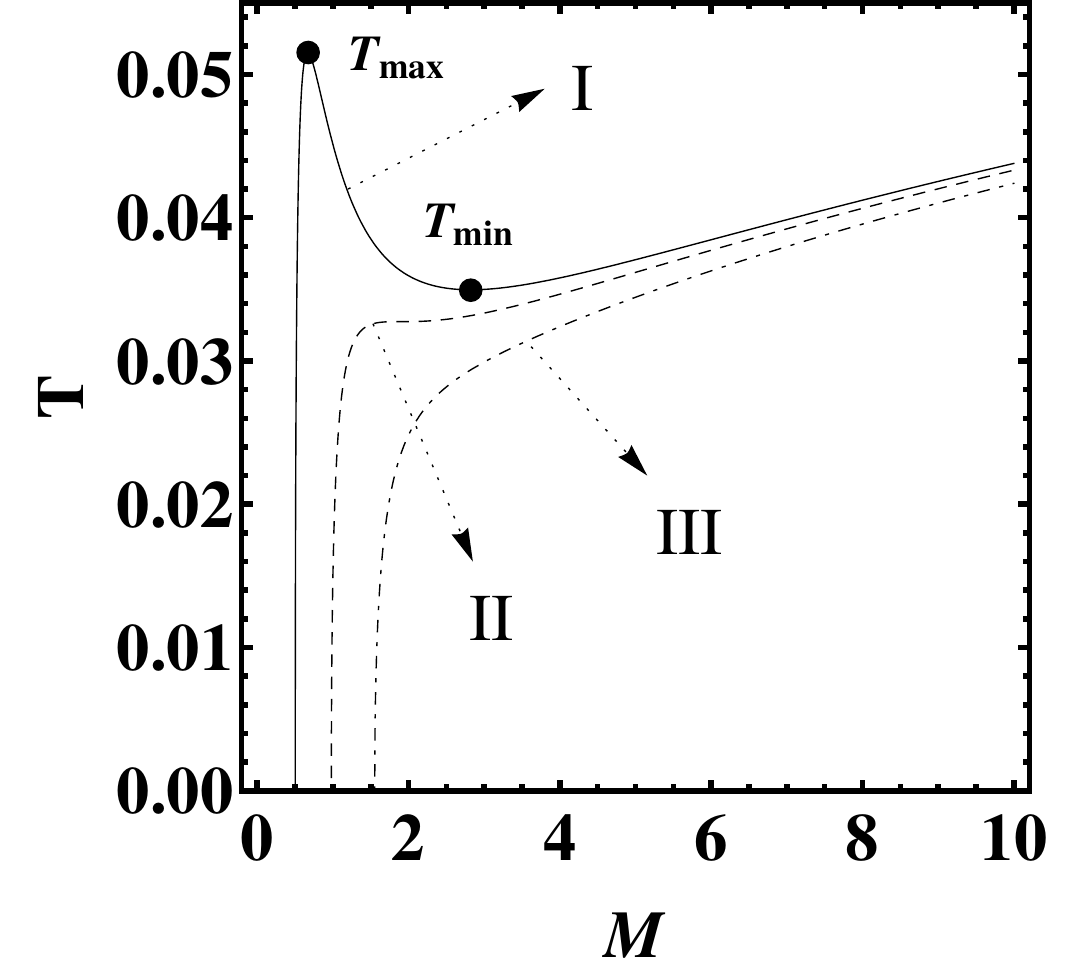}\,\,\,\,\,\,\,\,\,\,\,\,\,\,
		\includegraphics[width=0.40\textwidth]{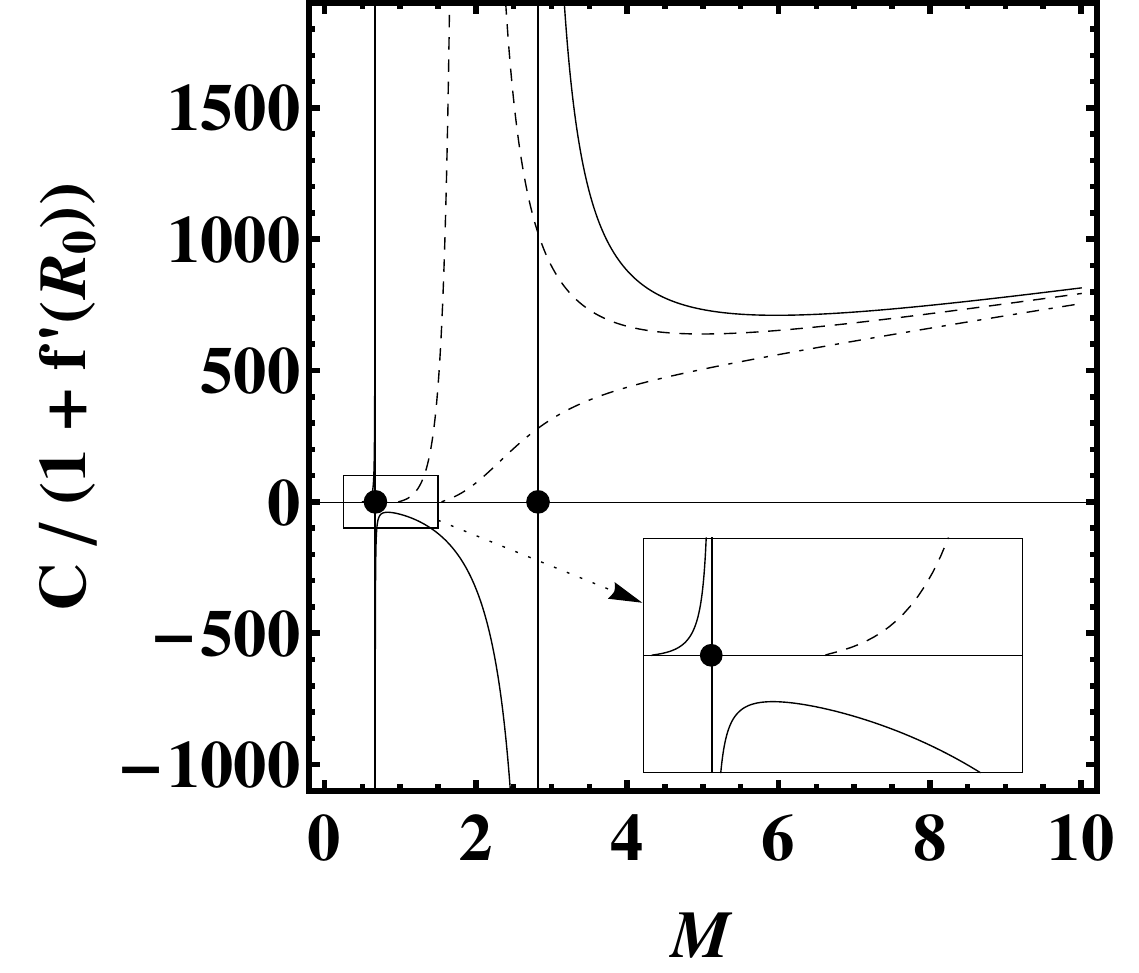}
\\[1cm]
		 \includegraphics[width=0.38\textwidth]{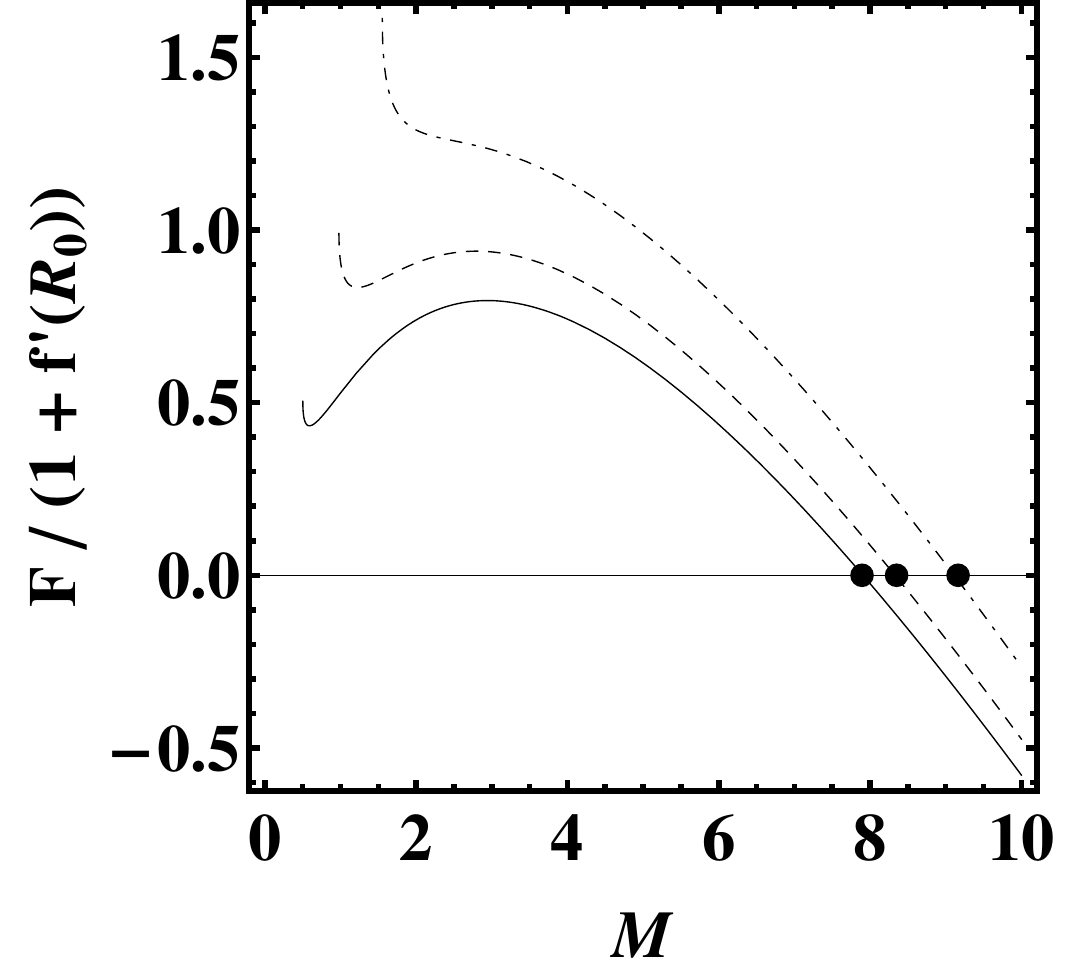}\,\,\,\,\,\,\,\,\,\,\,\,\,\,\,\,\,\,\,\,\,\,\,\,\,\,\,\,
		\includegraphics[width=0.348\textwidth]{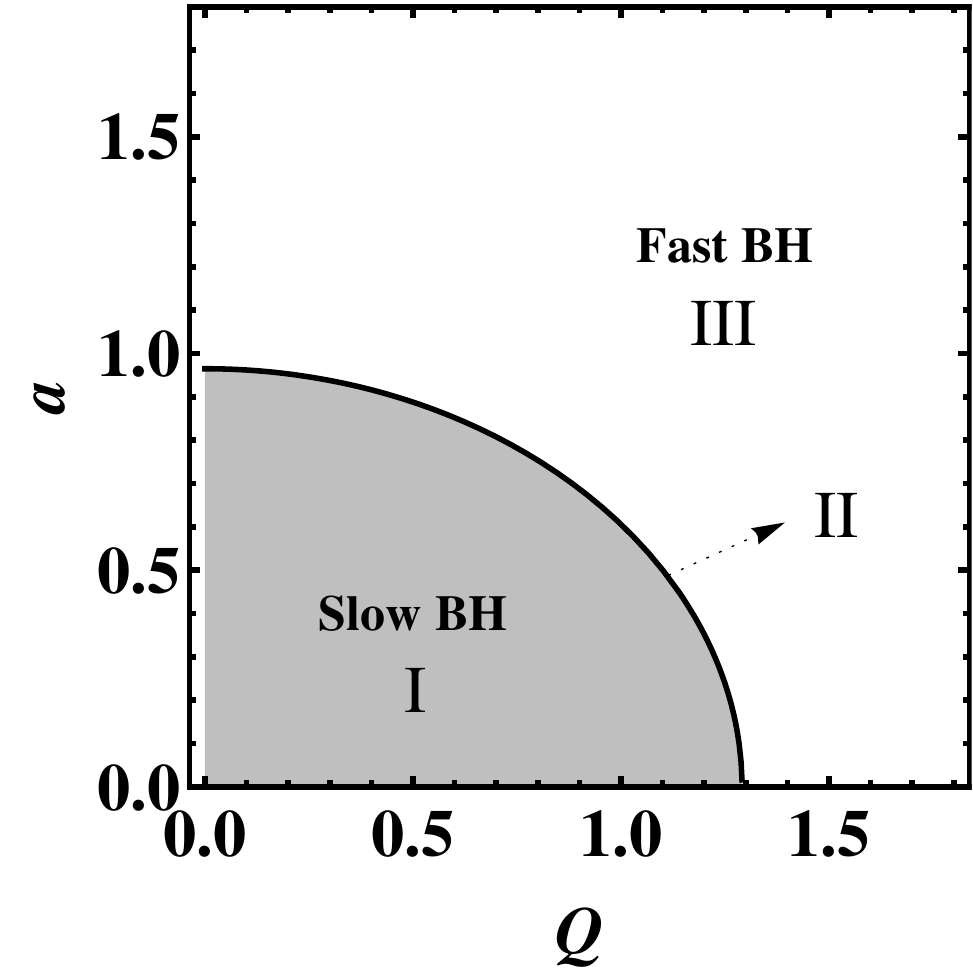}\,\,\,\,\,\,
		\caption{\footnotesize{For $R_0=-0.2$, we graphically display temperature (up to the left), heat capacity (up to the right), and the free energy (down to the left) of a BH as functions of the mass parameter $M$ for the cases: {\bf I)} $a=0.5$ y $Q=0$ : ``slow'' BH that shows a local maximum temperature $T_{max}$ and a local minimum temperature $T_{min}$ at the points where the heat capacity diverges, taking the latter negative values between $T_{max}$ y $T_{min}$. {\bf II)} $a\approx0.965$ y $Q=0$ : limit case where $T_{max}$ and $T_{min}$ merge, hence resulting on an inflection point in the temperature and an always positive heat capacity. {\bf III)} $a=1.5$ y $Q=0$ : ``fast'' BH with both temperature and heat capacity monotone growing (always positive too). It can be seen that all the configurations acquire a value $F<0$ from a certain value of $M$ onward, given by $r_{ext}^{\,\text{\it limit}}$. The values of $M^{min}$, with $T=0$ and $C=0$, correspond to an {\it extremal} BH. Down to the right, for $R_0=-0.2$ aswell, we display the regions in which BH behave as ``slow'' or ``fast'' BH.}}
	\label{fig:temperatura}
\end{figure*}

In Figure \ref{fig:temperatura} we have visualized the behavior of the temperature $T$, the free energy $F$ and the heat capacity $C$ of a BH for different values of mass $M$, with fixed $a$, $Q$ and $R_0$ values
\footnote{$R_0$ value is  given by the considered $f(R)$ model through equation \eqref{ec_curvatura_constante}}
. It can also be seen the
range of $a$ and $Q$ parameters values that provide {\it slow} or {\it fast}
BH for a constant value of the scalar curvature $R_0$.

Unlike Schwarzschild-$AdS$ BH case \cite{Hawking&Page}, Kerr-Newman-$AdS$ BH are allowed for any value of the temperature $T$, hence stability of each BH will be exclusively given by the corresponding values of heat capacity $C$ and free energy $F$ as functions of the $a$, $Q$, $M$ and $R_0$ parameters, that ultimately define the BH. However, for a set of fixed values of $a$, $Q$ and $R_0$, the mass parameter must be bigger than a minimum $M^{\text{\it min}}$ (characterized by $T=0$) to have BH configuration, otherwise radiation is the only possible equilibrium up to such a minimum mass. For bigger masses, we shall distinguish between the {\it fast} and the {\it slow} BH.
{\it Fast} BH, with bigger values of the spin and the electric charge than the {\it slow} ones, shows a heat capacity always positive and a positive free energy up to a $M=M^{\,\text{\it limit}}$ value, and negative onwards. Thus, this BH is unstable against tunneling decay into radiation for mass parameter values of $M<M^{\,\text{\it limit}}$. For $M>M^{\,\text{\it limit}}$, free energy becomes negative, therefore smaller than that of pure radiation, that will tend to collapse to the BH configuration in equilibrium with thermal radiation.

The second situation, i.e.,  the {\it slow} BH shows a more complex thermodynamics, being necessary to distinguish between four regions delimited by the mass parameter values: $M^{min}<M^{\textnormal {I}}<M^{\textnormal {II}}<M^{\,\text{\it limit}}$. For $M^{min}<M<M^{\textnormal {I}}$ and for $M^{\textnormal {II}}<M<M^{\,\text{\it limit}}$, both the heat capacity and the free energy are positive, which means that the BH is unstable to decay by {\it tunneling} into radiation. If $M^{\textnormal {I}}<M<M^{\textnormal {II}}$, the heat capacity becomes negative but free energy remains positive, being therefore unstable to decay into pure thermal radiation or to larger values of mass.
Finally, for $M>M^{\,\text{\it limit}}$ the heat capacity is positive whereas the free energy is now negative, thus tending pure radiation to tunnel to the BH configuration in equilibrium with thermal radiation.

It is mandatory to say that, although not quantitatively, the thermodynamical behavior of these $f(R)$ BH is qualitatively similar to that of GR \cite{Caldarelli}.

\section{Particular Examples}

In this section we will study some particular $f(R)$ models.
For each model we will firstly study the  range of parameters that allows the existence of Kerr-Newman BH.
Secondly, we will focus on the thermodynamical quantities that define BH stability depending again upon the model range of parameters \cite{Footnote3}.

For the sake of simplicity, let us introduce the dimensionless variables:
\begin{eqnarray}
\frac{r}{M}\rightarrow r,\,\,\,\,\frac{a}{M}\rightarrow a,\,\,\,\,\frac{Q}{M}\rightarrow Q,\,\,\,\,R_0\,M^2\rightarrow R_0,
\end{eqnarray}
where $M$ is the mass parameter, $a$ the spin parameter, $Q$ the electric charge parameter and $R_0$ is the scalar curvature obtained as a solution of equation \eqref{ec_escalar}. The considered models are:


\subsection{Model I: $f(R)\,=\,\alpha |R|^{\beta}$ }

\begin{figure*}[tp]
	\centering
		\includegraphics[width=0.40\textwidth]{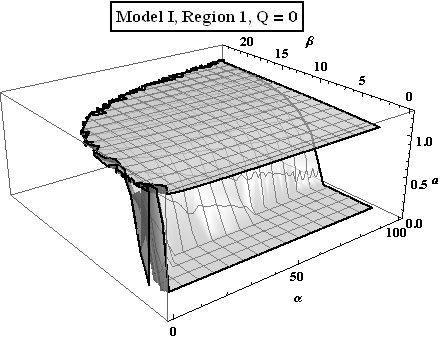}\,\,\,\,\,\,\,\,\,\,\,\,\,\,
		\includegraphics[width=0.40\textwidth]{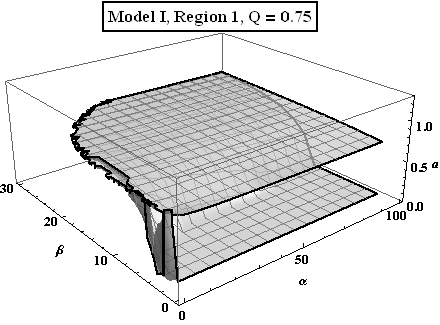}\\[0.5cm]
		\includegraphics[width=0.40\textwidth]{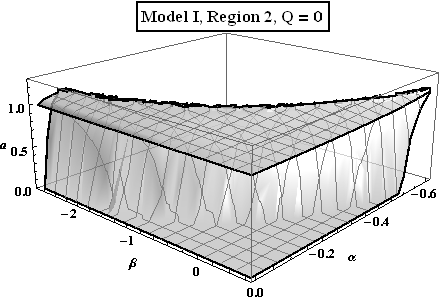}\,\,\,\,\,\,\,\,\,\,\,\,\,\,
		\includegraphics[width=0.40\textwidth]{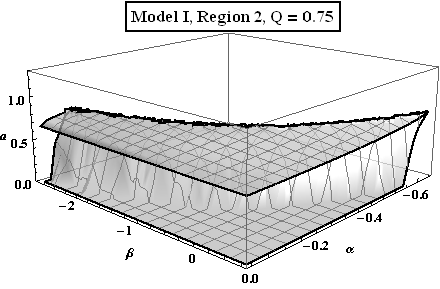}\\[0.5cm]
		\includegraphics[width=0.40\textwidth]{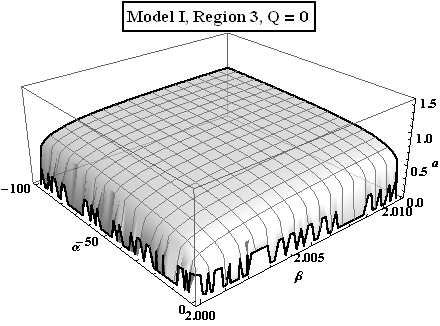}\,\,\,\,\,\,\,\,\,\,\,\,\,\,
		\includegraphics[width=0.40\textwidth]{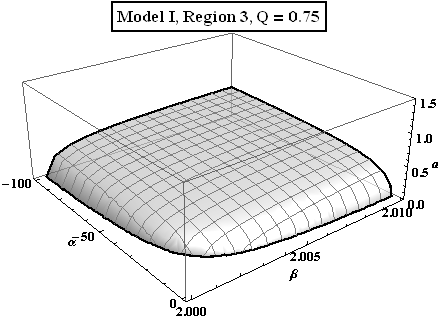}\\[0.5cm]
		\includegraphics[width=0.40\textwidth]{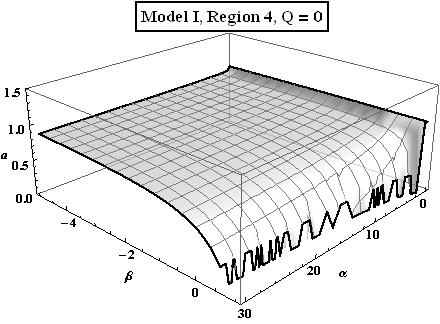}\,\,\,\,\,\,\,\,\,\,\,\,\,\,
		\includegraphics[width=0.40\textwidth]{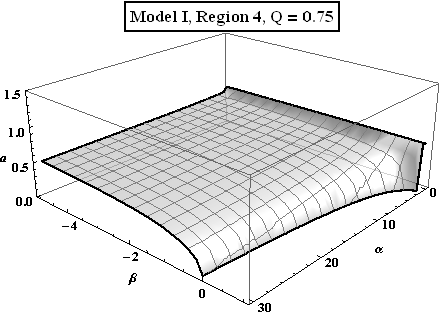}
		\caption{\footnotesize{
		\textbf{Model I}. Region 1: $\left\{ \alpha>0,\,\beta>2\right\}$, Region 2: $\left\{\alpha<0,\,\beta<1 \right\}$, Region 3: $\left\{ \alpha<0,\,\beta>2\right\}$ and Region 4: $\left\{ \alpha>0,\,\beta<1\right\}$. BH with a well defined horizon structure will only exist if they have a spin parameter below the upper surface $a_{max}$, and above a second surface $a_{min}$ (in case it exists, Regions 1 and 2 for this model) for certain values of $\alpha$ and $\beta$.
		}}
	\label{fig:modeloI}
\end{figure*}

This model has been widely studied because the $\alpha R^2$ term with $\alpha >0$ can account for the accelerated expansion of the Universe. This model can also explain the observed temperature anisotropies observed in the CMB, and could become a viable alternative to scalar field inflationary models; reheating after inflation would have its origin on the production of particles during the oscillation phase of the Ricci scalar \cite{Mijic}. By using expression \eqref{ec_curvatura_constante}, we obtain the following scalar curvature solutions:
\begin{eqnarray}
R_0^{\pm}=\pm\left[\frac{\pm 1}{(\beta-2)\,\alpha}\right]^{\frac{1}{\beta-1}},
\end{eqnarray}
where $R_0^+$ solution leads to a positive curvature, and $R_0^-$ to a negative one. The viability condition $1+f'(R_0^{\pm})>0$ restricts the range of parameters that define this $f(R)$ model to different regions depending on what solution we choose, $R_0^+$ or $R_0^-$; for $R_0^+$ we have Region 1 $\left\{ \alpha>0,\,\beta>2\right\}$ and Region 2 $\left\{\alpha<0,\,\beta<1 \right\}$, and for $R_0^-$, Region 3 $\left\{ \alpha<0,\,\beta>2\right\}$ and Region 4 $\left\{ \alpha>0,\,\beta<1\right\}$. In Figure \ref{fig:modeloI}, we show the range of the spin parameter $a$ for which BH are allowed, depending on the parameters $\alpha$ y $\beta$ and for certain values of the charge parameter $Q$. We graphically schematize in Figure \ref{fig:termoregiones} the possible thermodynamical configurations as functions of $\alpha$ and $\beta$, for those regions in which $R_0<0$.

\subsection{Model II: $f(R)\,=\,\pm|R|^{\alpha}\exp{\displaystyle \left(\frac{\beta}{R}\right)}-R$ }

\begin{figure*}[tp]
	\centering
		\includegraphics[width=0.40\textwidth]{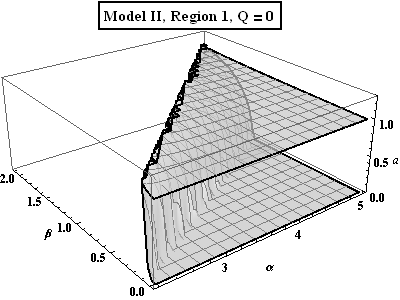}\,\,\,\,\,\,\,\,\,\,\,\,\,\,
		\includegraphics[width=0.40\textwidth]{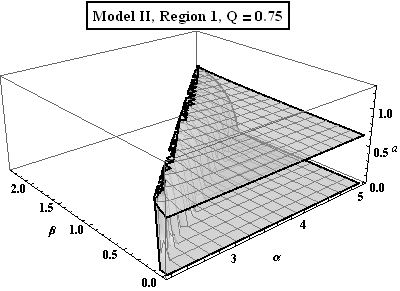}\\[0.5cm]
		\includegraphics[width=0.40\textwidth]{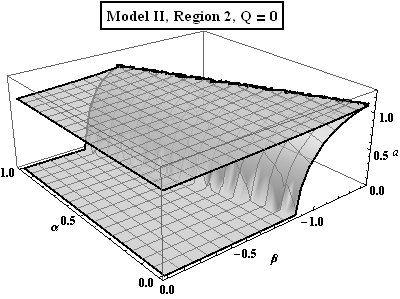}\,\,\,\,\,\,\,\,\,\,\,\,\,\,
		\includegraphics[width=0.40\textwidth]{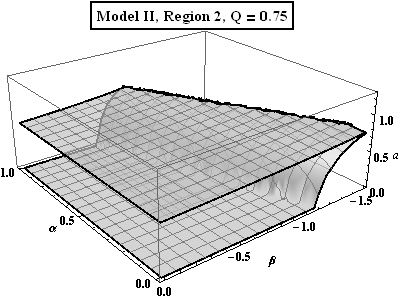}\\[0.5cm]
		\includegraphics[width=0.40\textwidth]{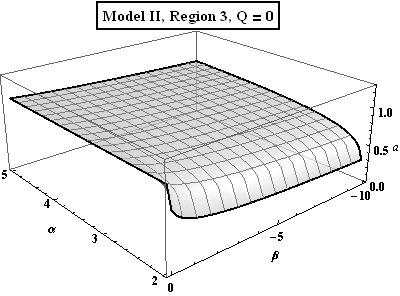}\,\,\,\,\,\,\,\,\,\,\,\,\,\,
		\includegraphics[width=0.40\textwidth]{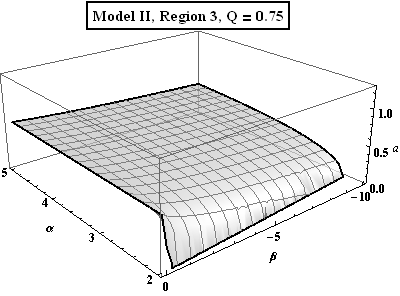}\\[0.5cm]
		\includegraphics[width=0.40\textwidth]{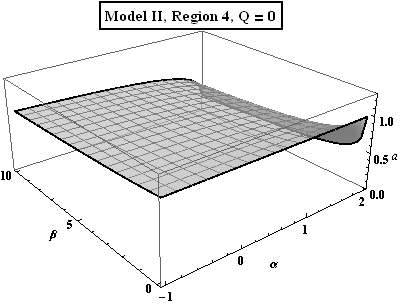}\,\,\,\,\,\,\,\,\,\,\,\,\,\,
		\includegraphics[width=0.40\textwidth]{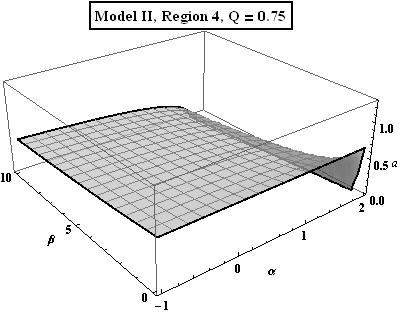}
		\caption{\footnotesize{
		\textbf{Model II}. Región 1: $\left\{ \alpha>2,\,\beta>0 \right\}$, Region 2: $\left\{ \alpha<2,\,\beta<0 \right\}$, Region 3: $\left\{ \alpha>2,\,\beta<0 \right\}$ and Region 4: $\left\{ \alpha<2,\,\beta>0 \right\}$. BH with a well defined horizon structure will only exist if they have a spin parameter below the upper surface $a_{max}$, and above a second surface $a_{min}$ (in case it exists, Regions 1 and 2 for this model) for certain values of $\alpha$ and $\beta$.		}}
	\label{fig:modeloII}
\end{figure*}

For this model and by using again expression \eqref{ec_curvatura_constante},
the scalar curvature, independently of the chosen sign, becomes:
\begin{eqnarray}
R_0=\frac{\beta}{\alpha-2}
\end{eqnarray}
However, the condition $1+f'(R_0)>0$ limits the theory $f(R)$ to different regions depending on what sign we decide to work with. If $R_{0}>0$, the theory is restricted to Region 1 $\left\{ \alpha>2,\,\beta>0 \right\}$ and Region 2 $\left\{ \alpha<2,\,\beta<0 \right\}$, for which $R_0$ takes positive values. If $R_{0}<0$, we restrict ourselves to Region 3 $\left\{ \alpha>2,\,\beta<0 \right\}$ and Region 4 $\left\{ \alpha<2,\,\beta>0 \right\}$. In Figure \ref{fig:modeloII} we show, just as before, the spin parameter $a$ values for which BH can exist in this model depending on the $\alpha$ and $\beta$ parameters defining the model. We graphically schematize in Figure \ref{fig:termoregiones} the possible thermodynamical configurations as functions of $\alpha$ and $\beta$, for those regions where $R_0<0$.

\subsection{Model III: $f(R)\,=\,R\,(\normalfont{\text{Log}}(\alpha\,R))^\beta-R$}

\begin{figure*}[tp]
	\centering
		\includegraphics[width=0.40\textwidth]{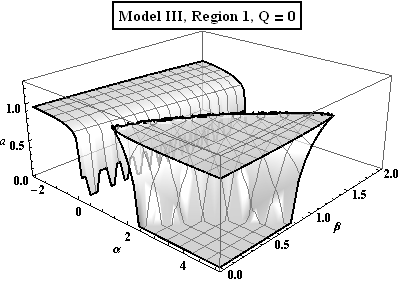}\,\,\,\,\,\,\,\,\,\,\,\,\,\,
		\includegraphics[width=0.40\textwidth]{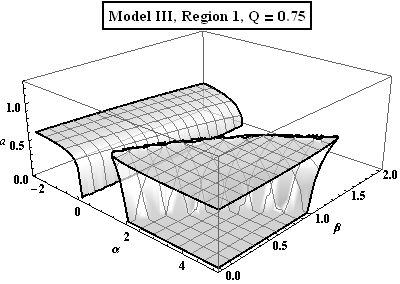}
		\caption{\footnotesize{
		\textbf{Model III}. Region 1: $\left\{ \alpha \in \mathbb R,\,\beta>0 \right\}$. BH with a well defined horizon structure will only exist if they have a spin parameter below the upper surface $a_{max}$, and above a second surface $a_{min}$ (that appears in this case for values of $\alpha>0$) for certain values of $\alpha$ and $\beta$.
		}}
	\label{fig:modeloIII}
\end{figure*}

The associated scalar curvature to this model is:
\begin{eqnarray}
R_0=\frac{1}{\alpha}\exp\,({\beta})
\end{eqnarray}
In this case the condition $1+f'(R_0)>0$ restricts us to work with Region 1 $\left\{ \alpha \in \mathbb R,\,\beta>0 \right\}$, where $R_0$ takes positive values for $\alpha>0$ and negatives for $\alpha<0$. In Figure \ref{fig:modeloIII} we graphically represent spin parameter $a$ values for which BH present their complete horizon structure, depending on the values of the parameters that define the model. We graphically schematize in Figure \ref{fig:termoregiones} the different possible thermodynamical configurations as functions of $\alpha$ and $\beta$, for those regions where $R_0<0$.

\subsection{Modelo IV: $f(R)\,=\,-\alpha\,\displaystyle \frac{\displaystyle \kappa\left(\frac{R}{\alpha}\right)^n}{\displaystyle 1+\beta\left(\frac{R}{\alpha}\right)^n}$ }

\begin{figure*}[tp]
	\centering
		\includegraphics[width=0.40\textwidth]{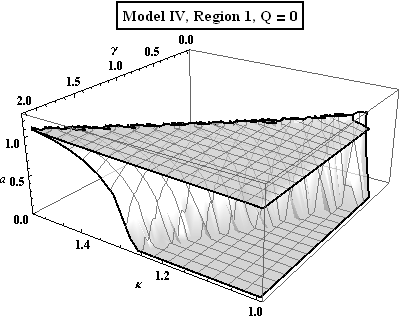}\,\,\,\,\,\,\,\,\,\,\,\,\,\,
		\includegraphics[width=0.40\textwidth]{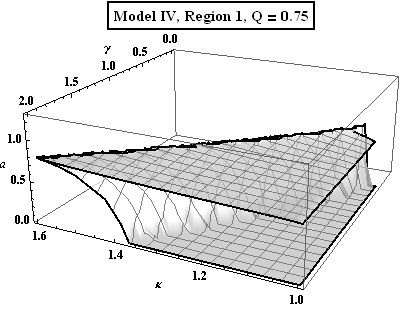}\\[0.5cm]
		\includegraphics[width=0.40\textwidth]{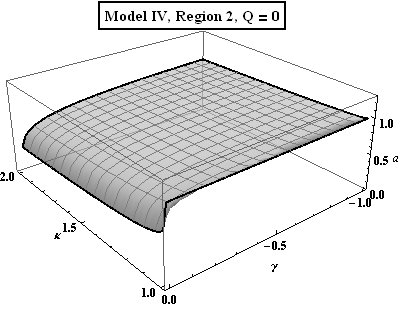}\,\,\,\,\,\,\,\,\,\,\,\,\,\,
		\includegraphics[width=0.40\textwidth]{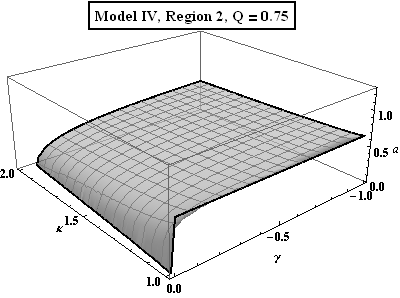}
		\caption{\footnotesize{
		\textbf{Model IV}. Region 1: $\left\{ \kappa>1,\,\gamma>0 \right\}$ and Region 2: $\left\{ \kappa>1,\,\gamma<0 \right\}$. BH with a well defined horizon structure will only exist if they have a spin parameter below the upper surface $a_{max}$, and above a second surface $a_{min}$ (in case it exists, only in Region 1 for this model) for certain values of $\kappa$ and $\gamma$.
		}}
	\label{fig:modeloIV}
\end{figure*}

\begin{figure*}[tp]
	\centering
		\includegraphics[width=0.40\textwidth]{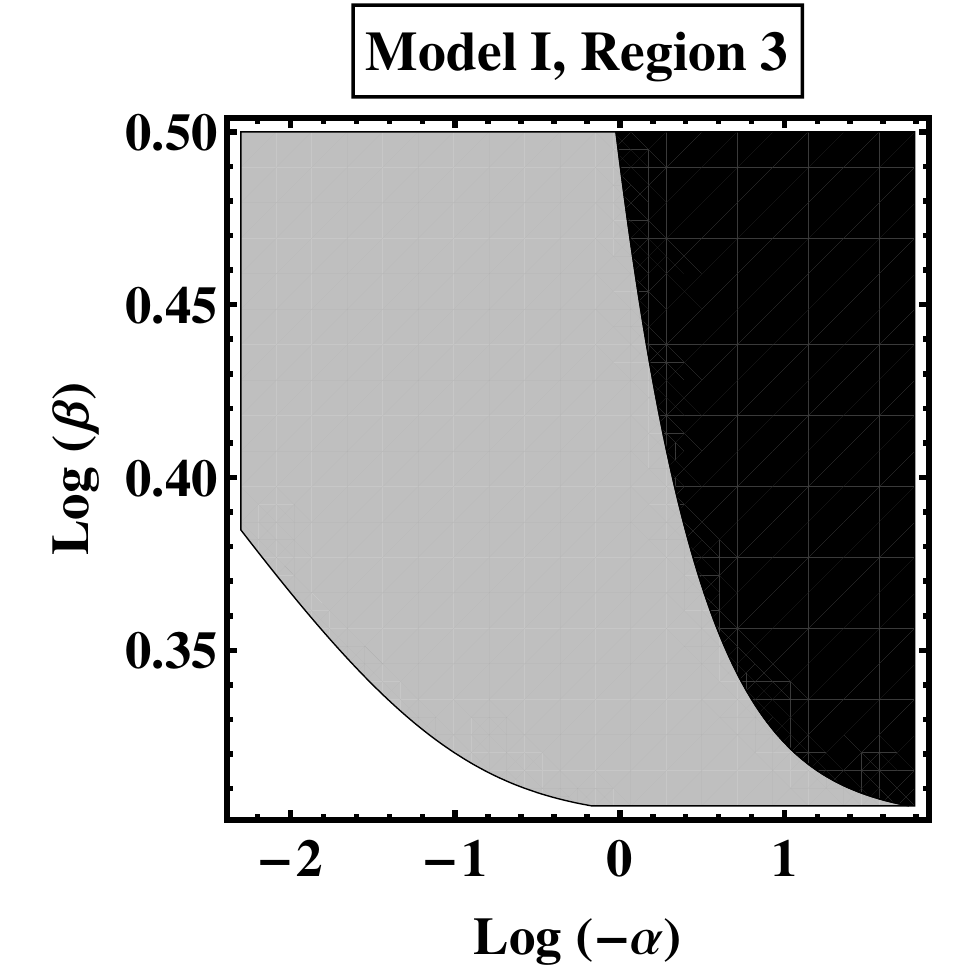}\,\,\,\,\,\,\,\,\,\,\,\,\,\,
		\includegraphics[width=0.40\textwidth]{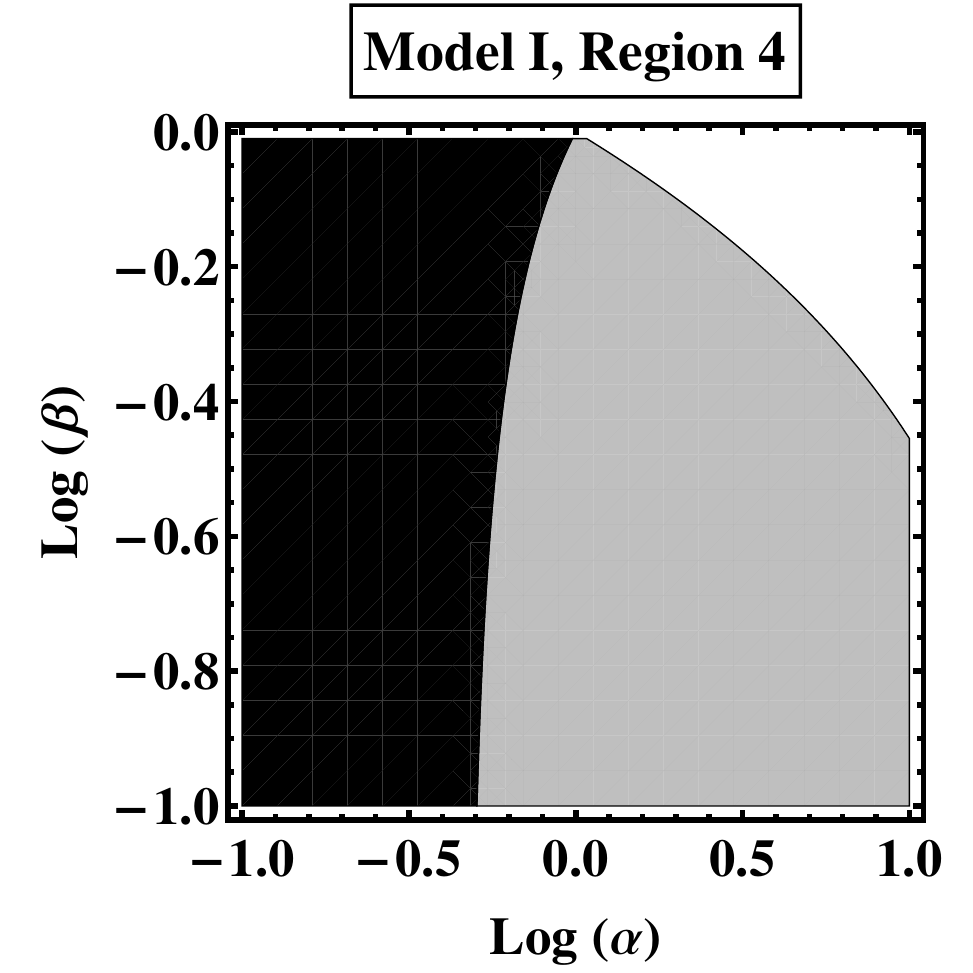}\\[0.8cm]						 			\includegraphics[width=0.40\textwidth]{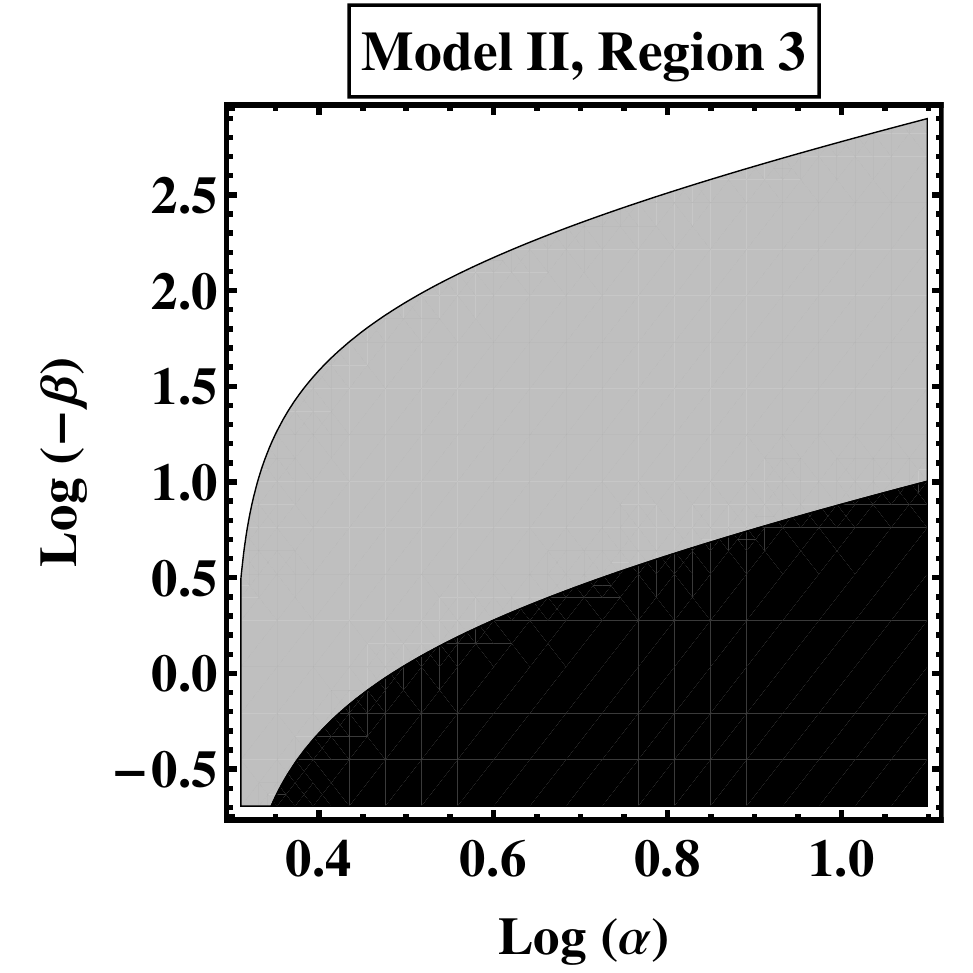}\,\,\,\,\,\,\,\,\,\,\,\,\,\,
		\includegraphics[width=0.40\textwidth]{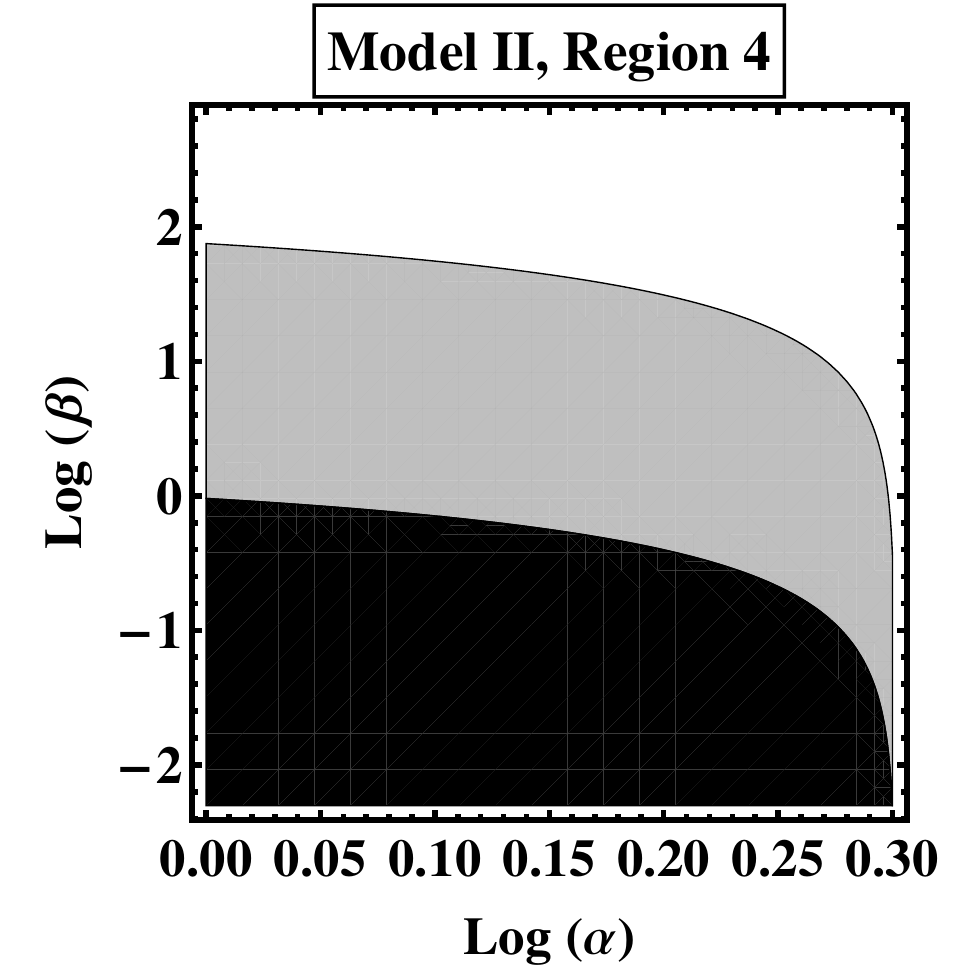}\\[0.8cm]
		\includegraphics[width=0.40\textwidth]{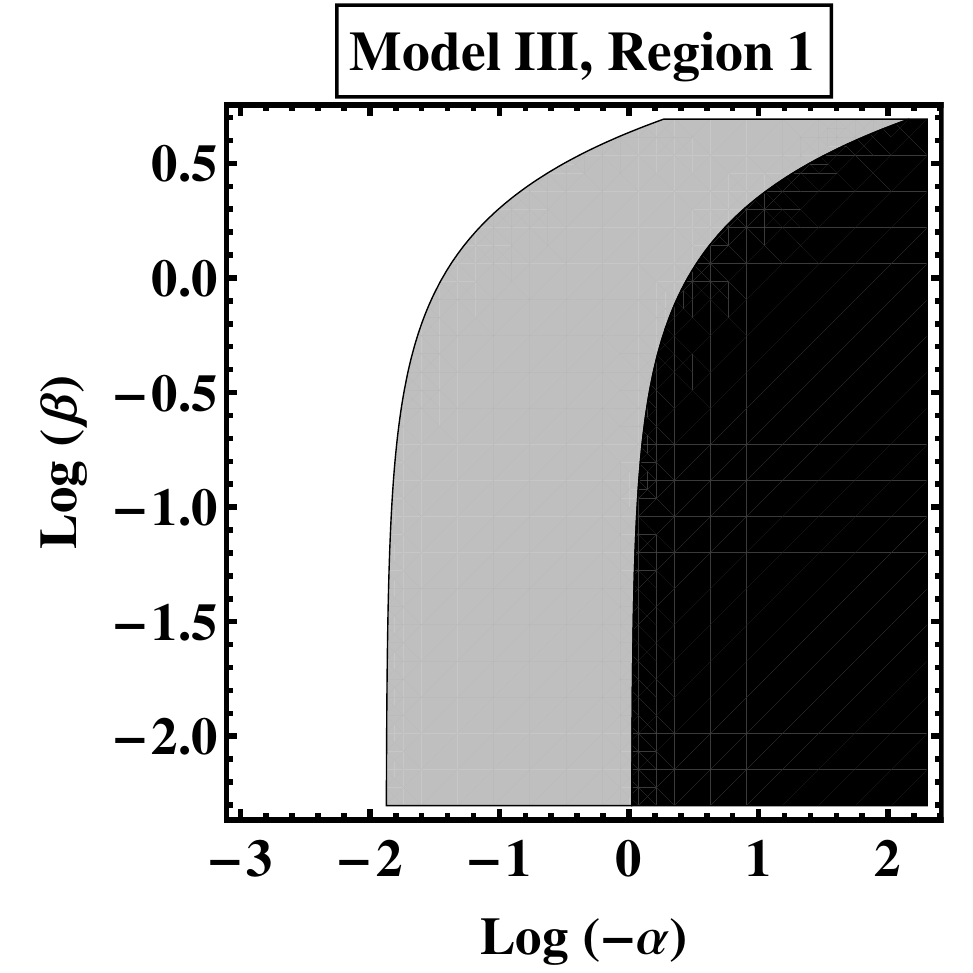}\,\,\,\,\,\,\,\,\,\,\,\,\,\,
		\includegraphics[width=0.40\textwidth]{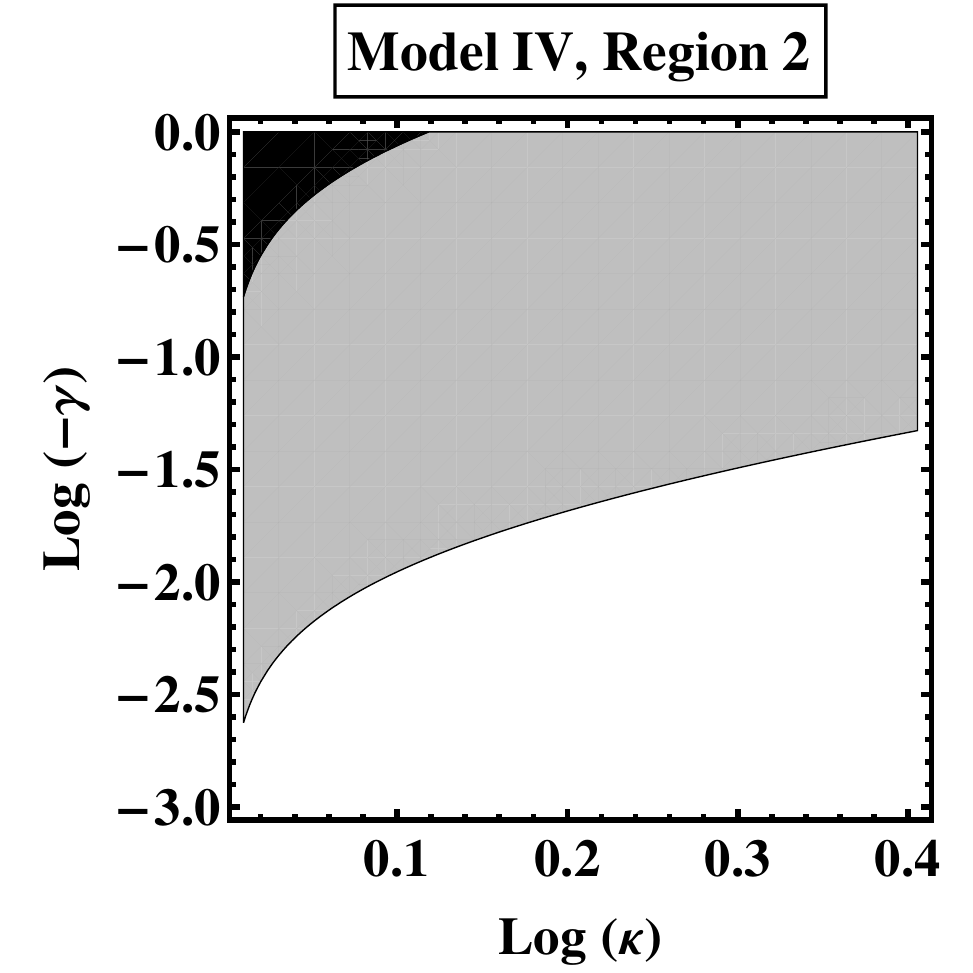}
		\caption{\footnotesize{Thermodynamical regions with negative scalar curvature $R_0<0$ of the models I, II, III and IV. For the sake of simplicity, we study a BH with the following parameter values: $M=1$, $a=0.4$ and $Q=0.2$. We distinguish between three different regions: {\bf  i)} $C<0$ and $F>0$, in black. {\bf ii)} $C>0$ and $F>0$, in gray. {\bf iii)} $C>0$ and $F<0$, in white.
		}}
	\label{fig:termoregiones}
\end{figure*}

This model has been proposed \cite{Hu&Sawicki2007} as cosmologically viable. For our study we will consider the case $n=1$, thus having a biparametric theory, as we can define $\gamma=\beta/\alpha$ and then obtain:
\begin{eqnarray}
f(R)\,=\,-\frac{\kappa\,R}{1+\gamma\,R}
\end{eqnarray}
Replacing the latter in \eqref{ec_curvatura_constante} we obtain two different values for the curvature:
\begin{eqnarray}
R_0^{\pm}=-\frac{1-\kappa}
{\gamma}\pm\sqrt{\frac{-\kappa\,(1-\kappa)}{\gamma^2}},
\end{eqnarray}
Keeping in mind that we have to satisfy $1+f'(R_0)>0$, $\kappa$ happens to be restricted to values $\kappa>1$.
On the other hand, computation of $1+f'(R_0^\pm)$ reveals that $R_0^+$ is only a valid solution for values of $\kappa$ and $\gamma$ in the Region 1: $\{ \kappa>1,\,\gamma>0\}$, and $R_0^-$ only in Region 2: $\{ \kappa>1,\,\gamma<0\}$,
being $R_0^{+}>0$ and $R_0^{-}<0$
in their respective regions. In Figure \ref{fig:modeloIV} we show the range of the spin parameter $a$ for which BH are allowed, depending on the parameters $\kappa$ and $\gamma$ for certain values of the charge $Q$. We graphically schematize in Figure \ref{fig:termoregiones} the different possible thermodynamical configurations as functions of $\kappa$ and $\gamma$, for those regions in which $R_0<0$.

\section{Conclusions}

In this work we have derived the metric tensor that describes a massive, charged, spinning object for $f(R)$ gravity in metric formalism. We found that it differs from that found by Carter \cite{Carter} by a multiplicative factor in the electric charge and a redefinition of vacuum scalar curvature.

Further study of the metric allowed us to describe the different astrophysical objects derived from the presence of different horizons: BH, {\it extremal} BH, {\it marginal extremal} BH, {\it naked singularities} and {\it naked extremal singularities}. Focusing on BH and their horizon structure, we have seen that these can only exist for values of the spin lower than a maximum value $a_{max}$, and that from a certain positive value of the curvature onward, only above a minimum value $a_{min}$.
We have also studied the thermodynamics of $AdS$-like BH (negative curvature solutions) by employing the Euclidean action method. It is observed that some quantities such as the mass, the energy or the entropy of these BH differ from those predicted in GR by a multiplicative factor $1+f'(R_0)$. This factor has to be positive in order to assure a positive mass and entropy for these kind of BH.

On the other hand, analysis of the behavior of the heat capacity of these BH reveals that we can distinguish between two kind of BH: {\it fast} and {\it slow}, showing the latter two phase transitions. We have also investigated the stability of the different possible configurations that arise from the values of the free energy and the heat capacity, observing that qualitatively the situation is similar to that described by Kerr-Newman-$AdS$ BH. Finally, we considered four $f(R)$ models and analyzed graphically the previous obtained results.
Experimental checks to test the validity of a particular $f(R)$ model might be done not only by studying astrophysical BH stability but also
if quantum gravity scale is near TeV: since LHC would be producing about one {\it microBH} per second \cite{microBH}, stability and thermodynamical properties of the produced BH might shed some light about the underlying theory of gravity. In this sense, let us remind that the relation between BH mass and temperature in $f(R)$ theories would differ from that predicted by GR.
%
%
%
%
\\
\\
{\bf Acknowledgments:} This work has been supported by MICINN (Spain) project numbers FIS 2008-01323, FPA 2008-00592 and Consolider-Ingenio MULTIDARK CSD2009-00064. AdlCD also acknowledges financial support from National Research Foundation (NRF, South Africa) and kind hospitality of UCM, Madrid while elaborating the manuscript.



\begin{thebibliography}{99}
\bibitem{Footnote1} Note that the metric \eqref{metrica} would be singular at $a^2=-12/R_0$.
\bibitem{Footnote2}  If for example we want to recover GR with a cosmological constant taking $f(R_0)=-2\Lambda$, this relation turns into:
$S={\cal A}_{H}/4$, i.e. the famous Bekenstein result \cite{Bekenstein}.
\bibitem{Footnote3} Let us remind at this stage that thermodynamical analysis is restricted to $R_0<0$ condition as explained in the beginning of Section IV.
\end{thebibliography}
\end{document}